\documentclass[prd,twocolumn,floatfix,nofootinbib]{revtex4}
\usepackage{graphicx}
\usepackage{epstopdf}
\usepackage{amssymb}
\usepackage{amsmath}
\usepackage{graphicx,subfigure,color,dcolumn,booktabs,bm}
\usepackage{longtable,lscape}
\usepackage{graphicx}
\usepackage{txfonts}
\usepackage{overpic}
\usepackage{indentfirst}
\usepackage{cases}
\usepackage{multirow}
\usepackage{ulem}
\usepackage[colorlinks,
citecolor=blue,
anchorcolor=red,
menucolor=red,
linkcolor=red,
filecolor=red,
runcolor=red,
urlcolor=blue,
frenchlinks=false]{hyperref}

\graphicspath{{Figures/}}
\allowdisplaybreaks
\DeclareRobustCommand{\revblue}[1]{%
  \textcolor{blue}{#1}%
}

\begin{document}
	
	\title{{
		A unified mass relation for conventional hadrons and fully heavy tetraquarks 
	}}
	
	\author{Wen-Nian Liu$^{1,2}$}
	\author{Duo-Jie Jia$^{3}$}
	\email[Corresponding author: ]{jiadj@nwnu.edu.cn}
	
	\author{Fu-Quan Dou$^{1,4}$}
	\email[Primary corresponding author: ]{doufq@nwnu.edu.cn}
	\affiliation{
		$^1$College of Physics and Electronic
		Engineering, Northwest Normal University, Lanzhou 730070, China\\
		$^2$Xinjiang Laboratory of Phase Transitions and Microstructures in Condensed Matters,College of Physical Science and Technology, Yili Normal University, Yining, Xinjiang, 835000, China\\
		$^3$ General Education Center,
		Qinghai Institute of Technology,
		Xining 810000, China\\
		$^4${Gansu} Provincial Research Center for Basic Disciplines of Quantum Physics, Lanzhou 730000, China }
	\date{\today}

	\begin{abstract}
We construct a unified mass relation for conventional hadrons and fully heavy tetraquarks based on a common reduced-mass function for effective chromoelectric interactions. Explicit color factors connect \(q\bar{q}\) and \(qq\) pairs, and fixed spin-weighted chromomagnetic contributions are subtracted. Using shared constituent quark masses and an additive constant, we jointly fit fourteen spin-averaged meson and baryon masses with a root-mean-square (RMS) residual of 4~MeV. The resulting saturating form links the nearly saturated light-flavor contribution with the increasingly attractive heavy-flavor behavior. In the common saturation limit, the leading chromoelectric contribution can be absorbed into effective constituent masses, while the residual pair term accounts for the flavor-dependent deviations. With the same parameter set, the model reproduces the conventional hadron spectra and predicts a fully charmed tetraquark tensor mass of \(6505(9)(9)\)~MeV, close to recent lattice QCD results, while the upper scalar state lies in the mass region of \(X(6600)\).
	\end{abstract}

	\maketitle
	
	\section{Introduction}
	\label{sec:introduction}
	
	In effective model descriptions of quantum chromodynamics (QCD)~\cite{DeRujula1975}, hadron masses can be organized as the sum of constituent quark masses, spin-independent contributions, and chromomagnetic contributions. The chromoelectric contribution forms the color-dependent part of the spin-independent term and helps set the overall mass scale. Treating this contribution consistently across different flavors and color configurations is a key step in connecting the descriptions of conventional hadrons and multiquark states.
	
	Potential models treat the chromoelectric Coulomb and confining interactions through spatial dynamics and describe meson and baryon spectra~\cite{Godfrey1985,Capstick1986,Brau2002}, while four-body calculations examine how mass ratios, color mixing, and the confining force affect multiquark binding~\cite{AderRichardTaxil1982,Zouzou1986,Vijande2013ColorMixing,Richard2018,Meng2025}. Recent screened-potential calculations have been extended to triply heavy tetraquark systems~\cite{Yang2026TriplyHeavy}, providing a dynamical reference for understanding chromoelectric effects. Complementarily, static effective approaches constrain mass relations using conventional hadron spectra. Methods based on mass relations and symmetry fix pair binding energies and estimate the masses of doubly heavy tetraquarks by combining color factors or heavy-quark symmetry~\cite{KarlinerRosner2017,EichtenQuigg2017,Zhang2026HADS}. The CMI model retains chromomagnetic terms explicitly to describe the color-spin correlations of multiquark systems~\cite{Jaffe1977} and configuration mixing~\cite{Hogaasen2006,Buccella2007}, while chromoelectric and other spin-independent contributions are usually absorbed into effective constituent masses~\cite{Stancu2010}. The reference-hadron approach further calibrates the chromomagnetic spectrum against known masses~\cite{Wu2019CMI,Cheng2020} and has been applied to hidden-flavor, doubly heavy, and singly heavy tetraquark systems~\cite{Li2024Hidden,Li2024Doubly,Man2026SinglyHeavy}. In addition, the chromoelectric contribution can be introduced explicitly as an independent pair term~\cite{Hogaasen2014Chromoelectric} or merged with the constituent masses into flavor-dependent effective quark-pair mass parameters~\cite{Weng2018,Weng2021,Weng2022,Guo2022,Guo2023}.
	
	On this basis, a further question arises: within a static effective framework, can a single parameter set describe the chromoelectric contribution across different flavors and color configurations while retaining the explicit two-body color factors that connect \(q\bar{q}\) and \(qq\) pairs? If so, its flavor dependence can be determined jointly from meson and baryon masses, thereby extending the mass relations constrained by conventional hadron spectra to multiquark systems.

	Simple constituent-quark models provide clues for developing such a unified description. In particular, additional attraction between heavy flavors relative to light-flavor reference systems was identified in Refs.~\cite{KR2014,KR2018} and incorporated into other model calculations~\cite{Zhang2021Bag,Yan2023Binding}. Using experimental meson masses, Ref.~\cite{KR2018Pwave} derived pair-binding parameters through quark-number and flavor conservation. Table~\ref{tab:meson_parameters_comparison} lists the binding parameters and constituent masses from the two static models of Refs.~\cite{Zhang2025Binding,KR2018Pwave}. Both sets show the same trend: the additional attraction increases from heavy--strange to heavy--heavy pairs. This trend guides our exploration of reduced-mass parametrizations.
	
	The reduced mass is the natural mass variable for two-body relative motion~\cite{QuiggRosner1979}. Guided by the color-Coulomb and confining interactions in the potential models discussed above, as well as by saturating and Cornell-type potential forms, we consider two parameterizations of the reduced-mass dependence. For each input set, we calculate $\mu_{ij}=m_i m_j/(m_i+m_j)$ using its own constituent masses and fit the same five binding terms with each form. The two parameterizations are
	\begin{equation}
		\label{eq:binding_fit_function}
		f_{\rm bind}(\mu)
		=a\mu+\frac{b}{c}(1-e^{-c/\mu})+d,
	\end{equation}
	and
	\begin{equation}
		\label{eq:binding_coulomb_inspired}
		g_{\rm bind}(\mu)=\alpha\mu+\frac{\beta}{\mu}+\gamma,
	\end{equation}
	where $a$ and $\alpha$ are dimensionless, $b$ and $\beta$ have dimensions of energy squared, and $c$, $d$, and $\gamma$ have dimensions of energy.
	
\begin{figure}[htbp]
		\centering
		\includegraphics[width=\linewidth]{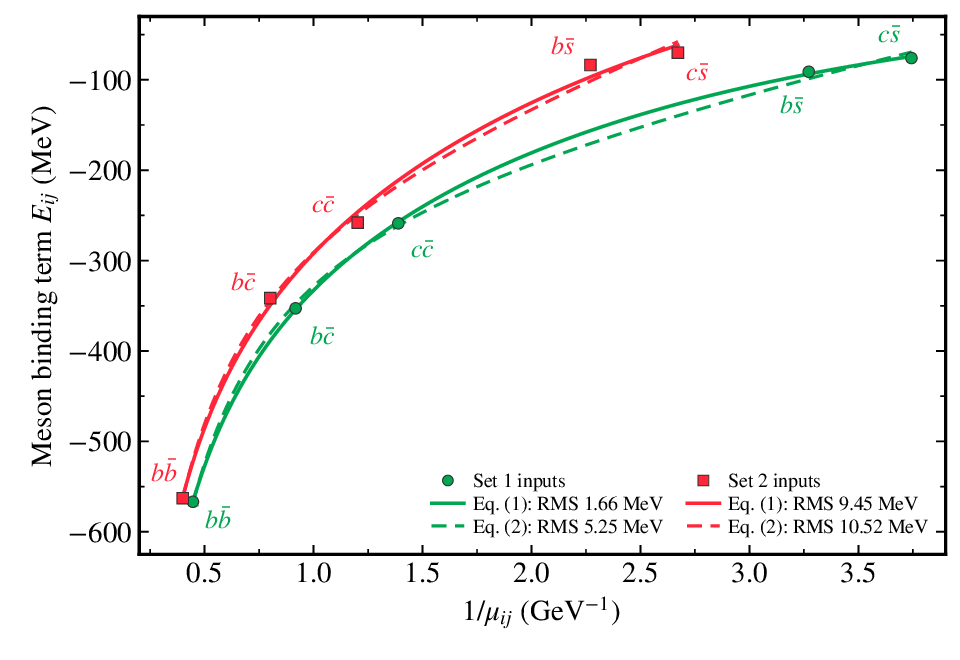}
		\caption{
			Color-singlet meson binding terms versus the inverse reduced mass. Green circles: Ref.~\cite{Zhang2025Binding}; red squares: Ref.~\cite{KR2018Pwave}. Solid and dashed curves correspond to Eqs.~\eqref{eq:binding_fit_function} and \eqref{eq:binding_coulomb_inspired}, respectively. Root-mean-square residuals (MeV) are given in the legend.}
		\label{fig:meson_binding_comparison}
	\end{figure}

	Both forms are intended only to capture the overall trend of the binding inputs. Fig.~\ref{fig:meson_binding_comparison} compares the fits. For Ref.~\cite{Zhang2025Binding}, Eqs.~\eqref{eq:binding_fit_function} and \eqref{eq:binding_coulomb_inspired} give root-mean-square residuals of 2 and 5 MeV, respectively. For Ref.~\cite{KR2018Pwave}, they give 9 and 11 MeV, respectively. The saturating form is slightly preferred. Because this comparison is exploratory, we do not use it to fix the final form. The final form is left to the subsequent joint meson--baryon fits and parameter-stability tests.
	
	\begin{table}[!t]
		\caption{Quark masses and color-singlet meson binding inputs (MeV); $n=u,d$. Attractive binding is negative.}
		\label{tab:meson_parameters_comparison}
		\centering
		\small
		\begin{ruledtabular}
			\begin{tabular}{lrr}
				Parameter
				& Ref.~\cite{Zhang2025Binding}
				& Ref.~\cite{KR2018Pwave} \\
				\hline
				$m_n$          & 230    & 310      \\
				$m_s$          & 328    & 483      \\
				$m_c$          & 1440   & 1663.300 \\
				$m_b$          & 4480   & 5003.800 \\
				\hline
				$B_{c\bar{s}}$ & $-76$  & $-70$  \\
				$B_{b\bar{s}}$ & $-91$  & $-84$  \\
				$B_{c\bar{c}}$ & $-259$ & $-258$ \\
				$B_{b\bar{c}}$ & $-353$ & $-342$ \\
				$B_{b\bar{b}}$ & $-567$ & $-563$ \\
			\end{tabular}
		\end{ruledtabular}
	\end{table}
	
	These exploratory fits indicate a functional relation between the additional heavy-quark binding energy, i.e., the average short-range chromoelectric contribution, and the reduced mass.
	
	The present work describes the flavor dependence of the effective chromoelectric contribution through a common reduced-mass function, linking the nearly saturated light-flavor contribution to the enhanced attraction in the heavy-flavor region, and connects the \(q\bar{q}\) and \(qq\) pairs through explicit two-body color factors. The parameters are jointly calibrated against the spin-averaged masses of mesons and baryons, and the fits and stability tests support the saturating form, with a root-mean-square residual of 4~MeV. After including the chromomagnetic interaction, the baryon spectrum is reconstructed with the same parameter set, and the predictions for multiply heavy baryons outside the calibration set are compared with lattice QCD results. The sum of the chromoelectric and chromomagnetic mass matrices is then diagonalized in bases that respect exchange symmetry, yielding the ground-state spectra of \(cc\bar{c}\bar{c}\), \(bb\bar{b}\bar{b}\), and \(bc\bar{b}\bar{c}\), and the fully charmed levels are compared with the mass and quantum numbers of \(X(6600)\) as well as with other model and lattice QCD results.
	
	The remainder of this paper is organized as follows. Sections~\ref{sec:framework}, \ref{sec:conventional}, and \ref{sec:tetraquarks} describe the theoretical framework and parameter calibration, conventional hadron spectra, and tetraquark spectra, respectively; Sec.~\ref{sec:summary} summarizes the paper. The bases, mass matrices, and consistency checks are provided in Appendix~\ref{app:basis} and Appendix~\ref{app:matrices}.

	\section{Theoretical Framework and Methods}
	\label{sec:framework}
	
	\subsection[Unified effective mass relation]{Unified effective mass relation}

	For the spin-independent two-body Hamiltonian in the center-of-mass frame
	\begin{equation*}
		H=\frac{\mathbf p^2}{2\mu}+V(r),
		\qquad
		\mu=\frac{m_i m_j}{m_i+m_j},
	\end{equation*}
	if the potential $V(r)$ is independent of the constituent masses, the ground-state energy $E_0$, excluding the constituent rest masses, satisfies~\cite{Feynman1939,QuiggRosner1979,Roncaglia1995}
	\begin{equation*}
		\frac{dE_0}{d(1/\mu)}
		=\frac{\langle\mathbf p^2\rangle}{2}>0,
		\qquad
		\frac{d^2E_0}{d(1/\mu)^2}\leq0.
	\end{equation*}
	That is, the energy rises with increasing inverse reduced mass while its slope gradually decreases, consistent with the overall trend in Fig.~\ref{fig:meson_binding_comparison}.
	
	Building on this, the specific functional form can be motivated by the scaling behavior of two interactions. On the one hand, a Coulomb interaction of fixed strength gives an energy scale proportional to $\mu$, with a characteristic radius proportional to $1/\mu$~\cite{QuiggRosner1979}; on the other hand, estimating at this reference scale a screened potential that approaches a constant at large distances~\cite{LiChao2009Screened} yields the saturating dependence $1-e^{-C/\mu}$. Combining these two aspects, we adopt the effective pair function
	\begin{equation}
		\label{eq:pair_function}
		F(\mu)
		=A_{\rm col}\mu
		+\frac{B_{\rm col}}{C}(1-e^{-C/\mu}),
	\end{equation}
	where $A_{\rm col}$ is dimensionless, $B_{\rm col}$ has dimensions of energy squared, and $C>0$ is the energy scale controlling the saturation transition. As a cross-check, we also consider the Cornell-inspired form $F_{\rm Cornell}(\mu)=A_{\rm Cornell}\mu+B_{\rm Cornell}/\mu$. The two forms share the same heavy-mass asymptotic structure; the main difference is that the saturating form approaches a finite value in the light-mass limit. The fits and stability tests with the same inputs presented below favor the saturating form, so all subsequent spectrum calculations use its calibrated parameters.
	
Keeping the two-body color factors explicit, we write the unified effective mass operator as
\begin{equation}
	\label{eq:unified_mass_operator}
	\begin{aligned}
		\widehat{\mathcal M}_H
		&=\left(\sum_{i=1}^{N_H}m_i+D\right)\mathbf I
		+\sum_{i<j}F(\mu_{ij})\,\mathbf T_i\!\cdot\!\mathbf T_j\\
		&\quad-4\sum_{i<j}C_{ij}
		(\mathbf T_i\!\cdot\!\mathbf T_j)
		(\boldsymbol\sigma_i\!\cdot\!\boldsymbol\sigma_j).
	\end{aligned}
\end{equation}
Here \(N_H\) is the total number of quarks and antiquarks, \(\mathbf I\) is the identity in color--spin space, \(\boldsymbol\sigma_i\) is the vector of Pauli matrices acting on constituent \(i\), and \(C_{ij}\) denotes the effective chromomagnetic couplings (fixed inputs rather than fit parameters; see Sec.~\ref{sec:cmi_couplings}).
Mesons and baryons share the constituent masses and the parameters \(A_{\rm col},B_{\rm col},C,D\),
with equal masses for quarks and antiquarks of the same flavor.
The constant \(D\) represents an overall spin-independent hadron-level mass shift, counted once per hadron,
consistent with the treatment of hadron masses in some potential models~\cite{SilvestreBrac2003,Brau2002}.
With \(\mathbf T_q=\boldsymbol\lambda/2\) and \(\mathbf T_{\bar q}=-\boldsymbol\lambda^*/2\),
where \(\boldsymbol\lambda\) denotes the Gell-Mann matrices,
the pair color factors are \(-4/3\) for a color-singlet meson and \(-2/3\) for each pair in a color-singlet baryon.
This explicit color structure connects with the chromoelectric terms of extended CMI models~\cite{Weng2018,Weng2021}
and is motivated by approximate Casimir scaling at intermediate distances~\cite{Bali2000,Deldar2000}.
The calculated masses are the eigenvalues in the color-singlet space satisfying the exchange symmetries,
namely \(\widehat{\mathcal M}_H|\Psi_{H,k}\rangle
=M_{H,k}^{\rm cal}|\Psi_{H,k}\rangle\),
where \(|\Psi_{H,k}\rangle\) denotes the corresponding color--spin eigenstate.
	
	For mesons and baryons, the spin-independent part of Eq.~\eqref{eq:unified_mass_operator} gives
	\begin{equation}
		\label{eq:spin_average_mass}
		M_{{\rm SI},H}
		=\sum_{i=1}^{N_H}m_i
		+\sum_{i<j}
		\langle\mathbf T_i\!\cdot\!\mathbf T_j\rangle_H
		F(\mu_{ij})+D.
	\end{equation}
	For a meson, the pair contribution and the constant term sum to \(-(4/3)F(\mu)+D\), which has the same functional form as Eq.~\eqref{eq:binding_fit_function}. The binding-energy interpolations in Fig.~\ref{fig:meson_binding_comparison} use the constituent masses of the respective references and are independent of the joint calibration here.
	
	The calibration uses spin-averaged conventional-hadron masses and retains their averaged hyperfine contributions~\cite{DeRujula1975,QuiggRosner1979,Godfrey1985,Capstick1986}. For a multiplet \(\mathcal S_H\), we define
	\begin{equation*}
		\bar M_H=
		\frac{\sum_{k\in\mathcal S_H}(2J_k+1)M_{H,k}}
		{\sum_{k\in\mathcal S_H}(2J_k+1)},
	\end{equation*}
	where \(J_k\) is the total spin and \(2J_k+1\) is the magnetic degeneracy. The corresponding spin-weighted chromomagnetic contribution is
	\begin{equation}
		\label{eq:cmi_average}
		\bar E_H=
		\frac{\sum_{k\in\mathcal S_H}(2J_k+1)E_{H,k}}
		{\sum_{k\in\mathcal S_H}(2J_k+1)},
	\end{equation}
	with \(E_{H,k}\) the chromomagnetic eigenvalue at fixed couplings. Hence \(\bar M_H^{\rm fit}=M_{{\rm SI},H}+\bar E_H\), and the spin-independent reference mass used in the fit is \(M_{{\rm SI},H}^{\rm ref}=\bar M_H^{\rm ref}-\bar E_H\), where the superscripts ``fit'' and ``ref'' denote fitted and reference values, respectively. The inputs and calibration are described in the next subsection, and the chromomagnetic couplings and individual-state reconstruction are given in Sec.~\ref{sec:cmi_couplings}.
	
	\subsection{14-point calibration and stability tests}
	\label{sec:calibration}
	
We determine the eight parameters of the spin-independent part of Eq.~\eqref{eq:unified_mass_operator} by an unweighted least-squares fit, minimizing
\begin{equation}
	\chi^2=\sum_{a=1}^{14}
	\left[
	M_{{\rm SI},a}
	-\left(\bar M_a^{\rm ref}-\bar E_a^{\rm CMI}\right)
	\right]^2.
\end{equation}
Here \(a\) labels the fourteen reference spin-averaged masses, and \(\bar E_a^{\rm CMI}\) is the fixed average defined in Eq.~\eqref{eq:cmi_average}, evaluated with the couplings of Sec.~\ref{sec:cmi_couplings} and held fixed in the fit.
	
	The fit uses fourteen inputs in total (hereafter the 14-point fit): seven mesons, $n\bar s$, $c\bar n$, $c\bar s$, $c\bar c$, $b\bar s$, $b\bar c$, and $b\bar b$, and seven baryons, $nnn$, $nns$, $nsc$, $ssc$, $nnb$, $nsb$, and $scc$. This input set covers all ten quark-pair flavor combinations and includes light, singly charmed, singly bottom, and doubly charmed baryons. All inputs are experimental reference values, or reference values constructed from experimental masses together with the hyperfine-splitting relations; lattice masses are used for comparison only. The fitted parameters are listed in Table~\ref{tab:stable14_parameters}.
	
	\begin{table}[!t]
		\caption{\label{tab:stable14_parameters}
			{
				Parameters from the equal-weight fit to seven meson and seven baryon
				inputs, with fixed CMI contributions subtracted and a common $D$.
			}
		}
		\centering
		\begin{ruledtabular}
			\begin{tabular}{lr}
				Parameter & Value \\
				\hline
				$m_n$ (GeV)
				& {{{0.234}}} \\
				$m_s$ (GeV)
				& {{{0.439}}} \\
				$m_c$ (GeV)
				& {{{1.645}}} \\
				$m_b$ (GeV)
				& {{{5.002}}} \\
				$A_{\rm col}$
				& {{{0.097}}} \\
				$B_{\rm col}$ (GeV$^2$)
				& {{{-0.162}}} \\
				$C$ (GeV)
				& {{{0.439}}} \\
				$D$ (GeV)
				& {{{-0.316}}} \\
			\end{tabular}
		\end{ruledtabular}
	\end{table}
	
	To test stability, we increase the number of inputs step by step from eleven to seventeen. Starting from the 14-point input set, we first remove $c\bar s$, $ssc$, and $nsb$, and then successively add $nsb$, $c\bar s$, $ssc$, $n\bar n$, $ncc$, and $b\bar n$. Each resulting input set covers all meson flavor sectors and all four baryon sectors, and the numbers of mesons and baryons differ by no more than one. For each input set, all eight parameters are refitted with the same weights and fixed couplings. Table~\ref{tab:stable14_parameter_stability} lists the percentage changes of the parameters relative to their 14-point values,
	$\epsilon_p^{(N)}=100|p^{(N)}-p^{(14)}|/|p^{(14)}|$,
	together with the root-mean-square residual
	\begin{equation}
		{\rm RMS}_N=
		\left[\frac1N\sum_{a\in S_N}
		(\bar M_a^{\rm fit}-\bar M_a^{\rm ref})^2\right]^{1/2},
	\end{equation}
	where $N$ is the number of inputs and $S_N$ the corresponding index set. Table~\ref{tab:stable14_parameter_stability} shows that all eight parameters remain within ${1.699}\%$ of their 14-point values, with maximum changes of only ${0.162}\%$ and ${0.080}\%$ for $m_c$ and $m_b$, respectively, indicating good fit stability. Consequently, the 14-point fit is used for all subsequent spectrum calculations.

	\begin{table*}[!t]
		\caption{\label{tab:stable14_parameter_stability}
			{
				Parameter changes along the eleven- to seventeen-point sequence:
				$\epsilon_p^{(N)}=100|p^{(N)}-p^{(14)}|/|p^{(14)}|$ (\%).
				The last row gives the sample RMS in MeV;
				$N_M$ and $N_B$ count meson and baryon inputs.
				Starting from eleven points, we add $nsb$, $c\bar s$, $ssc$,
				$n\bar n$, $ncc$, and $b\bar n$ in order.
				The 14-point RMS is 4 MeV.
			}
		}
		\centering
		\begin{ruledtabular}
			\begin{tabular}{lrrrrrrr}
				Parameter & $N=11$ & $N=12$ & $N=13$ & $N=14$ & $N=15$ & $N=16$ & $N=17$ \\
				\hline
				$N_M+N_B$ & $6+5$ & $6+6$ & $7+6$ & $7+7$ & $8+7$ & $8+8$ & $9+8$ \\
				\hline
				$m_n$ & {{0.187}} & {{0.015}} & {{0.005}} & {0.000} & {{0.084}} & {{0.549}} & {{0.681}} \\
				$m_s$ & {{0.423}} & {{0.013}} & {0.044} & {0.000} & {{0.205}} & {{0.445}} & {{0.500}} \\
				$m_c$ & {{0.162}} & {{0.036}} & {0.023} & {0.000} & {{0.081}} & {0.006} & {{0.001}} \\
				$m_b$ & {{0.080}} & {{0.015}} & {0.010} & {0.000} & {{0.031}} & {{0.018}} & {{0.018}} \\
				$A_{\rm col}$ & {{1.699}} & {{0.452}} & {{0.362}} & {0.000} & {{0.396}} & {{0.587}} & {{0.482}} \\
				$B_{\rm col}$ & {{1.493}} & {{0.577}} & {{0.447}} & {0.000} & {{0.674}} & {{0.038}} & {0.019} \\
				$C$ & {{1.673}} & {{0.602}} & {{0.461}} & {0.000} & {{0.903}} & {{0.618}} & {{0.749}} \\
				$D$ & {{0.393}} & {{0.112}} & {{0.142}} & {0.000} & {{0.388}} & {{0.069}} & {{0.025}} \\
				\hline
				RMS (MeV) & {{4.45}} & {{4.31}} & {{4.14}} & {{4.00}} & {{3.88}} & {{3.89}} & {{3.78}} \\
			\end{tabular}
		\end{ruledtabular}
	\end{table*}
	
	To assess the predictive power of the fit, we first perform leave-one-out refits over all inputs and obtain a predictive RMS residual of 19 MeV, which mainly reflects the effect of omitting the $b\bar b$ input. Because $b\bar b$ probes the deeply bound region at the largest reduced mass and places an important constraint on the heavy-flavor parameters, we retain it as a calibration anchor. With this anchor fixed, we perform leave-one-out tests on the remaining thirteen inputs and obtain a predictive RMS residual of 9 MeV and a maximum absolute residual of 18 MeV. Accordingly, under this calibration condition, we take $\pm9$ MeV as a typical empirical prediction scale for the spin-independent masses and use the maximum absolute residual to characterize the largest deviation in these tests.
	
	The experimental baryon averages use the same charge-state averages as the individual-state comparisons in Table~\ref{tab:stable14_baryon_spectrum_exp} and Table~\ref{tab:stable14_heavy_baryon_spectrum}~\cite{PDG2026}, with the 14 mass inputs and fixed CMI couplings left unchanged. In particular, the $b\bar c$ input includes the observed $B_c^*$ mass~\cite{ATLAS2026BcStar}, while the $s\bar s$ comparison uses the unmixed auxiliary state $\eta_s$~\cite{Zhang2025Binding}.
	
	For $ncc$ and $scc$, the measured spin-$1/2$ masses~\cite{LHCb2017Xi,LHCb2026OmegaCC}, combined with the fixed-coupling splitting relation $M_{3/2}-M_{1/2}=16C_{qc}$, give
	\begin{equation}
		\label{eq:hf_assisted_centroid}
		\bar M_{qcc}^{\rm ref}
		=M_{1/2}^{\rm exp}+\frac{32}{3}C_{qc},\qquad q=n,s.
	\end{equation}
	The $scc$ reference enters the 14-point fit, whereas $ncc$ is used for comparison only and is added at $N=16$ in the stability test; both are labeled Exp.+HF. Subtracting $\bar E_{qcc}^{\rm CMI}=8C_{cc}/3$ from the above expression yields the corresponding spin-independent reference mass. Here $C_{ij}$ is the fixed chromomagnetic coupling for quark pair $ij$,
	while $C$ is the crossover scale in $F(\mu)$; the two are distinct quantities.
	
	Reference spin-averaged masses and individual-state experimental masses are tabulated separately, and the lattice results~\cite{Brown2014Baryons} are used for comparison only. The spin weights are those defined above, with no additional isospin multiplicity for $nnn$. For $nns$, $nnc$, and $nnb$, we average over the $\Sigma$-type partners, with the corresponding $\Lambda$-type states sharing the same spin-independent mass; for $nsc$ and $nsb$, we use $\Xi'_Q$ and $\Xi_Q^*$ ($Q=c,b$); and for $ncb$ and $scb$, both spin-$1/2$ states and the spin-$3/2$ state are included.
	
	The resulting CMI averages are $8C_{nn}/3$ for $nnn$, $nns$, $nnc$, and $nnb$; $8C_{ss}/3$ for $nss$, $ssc$, and $ssb$; $8C_{cc}/3$ for $ncc$, $scc$, and $ccb$; and $8C_{bb}/3$ for $nbb$, $sbb$, and $cbb$. For $nsQ$, the higher of the two spin-$1/2$ CMI eigenvalues and the spin-$3/2$ eigenvalue are averaged with weights $1:2$. The complete $ncb$ and $scb$ multiplets have vanishing CMI averages.
	
	Table~\ref{tab:stable14_meson_averages} and Table~\ref{tab:stable14_baryon_averages} compare the reference and calculated values, with $\Delta\bar M=\bar M^{\rm fit}-\bar M^{\rm ref}$ and asterisks marking the fit inputs. The joint RMS residual is 4 MeV, as are the RMS residuals of the meson and baryon sectors separately. For references determined directly from experimental data, the largest deviations are 8 MeV for mesons and 10 MeV for baryons; the four unfitted entries have an RMS of 6 MeV. The auxiliary $s\bar s$ results and the Exp.+HF results are listed separately.
	
	\begin{table*}[!t]
		\caption{\label{tab:stable14_meson_averages}
			Meson spin-averaged masses and
			14-point fit results (MeV).
			Here $\bar M=(M_{0^-}+3M_{1^-})/4$ and
			$\Delta\bar M=\bar M^{\rm fit}-\bar M^{\rm ref}$.
			Asterisks mark fit inputs;
			the $s\bar s$ reference uses $\eta_s$.
		}
		\centering
		\renewcommand{\arraystretch}{1.10}
		\begin{ruledtabular}
			\begin{tabular}{llrrr}
				Flavor & States $(0^-,1^-)$
				& $\bar M^{\rm ref}$
				& $\bar M^{\rm fit}$ & $\Delta\bar M$ \\
				\hline
				$n\bar n$ & $\pi,\rho$
				& 616 & 618 & $+2$ \\
				$n\bar s^{\,*}$ & $K,K^*$
				& 794 & 802 & $+8$ \\
				$s\bar s$ & $\eta_s,\phi$
				& 937 & 959 & $+22$ \\
				$c\bar n^{\,*}$ & $D,D^*$
				& 1973 & 1971 & $-2$ \\
				$c\bar s^{\,*}$ & $D_s,D_s^*$
				& 2076 & 2076 & $0$ \\
				$c\bar c^{\,*}$ & $\eta_c,J/\psi$
				& 3069 & 3071 & $+2$ \\
				$b\bar n$ & $B,B^*$
				& 5313 & 5315 & $+1$ \\
				$b\bar s^{\,*}$ & $B_s,B_s^*$
				& 5403 & 5400 & $-3$ \\
				$b\bar c^{\,*}$ & $B_c,B_c^*$
				& 6323 & 6318 & $-5$ \\
				$b\bar b^{\,*}$ & $\eta_b,\Upsilon(1S)$
				& 9445 & 9446 & $+1$ \\
			\end{tabular}
		\end{ruledtabular}
	\end{table*}
	
	\begin{figure}[htbp]
		\centering
		\includegraphics[width=0.43\textwidth]{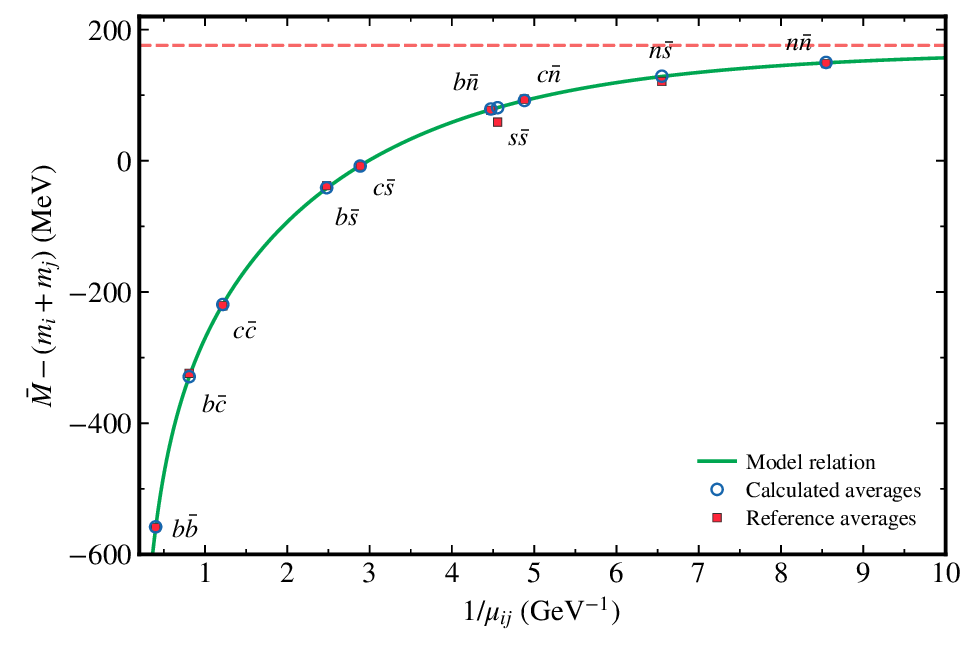}
		\caption{\label{fig:meson_effective_mass_contribution}
		Meson contributions $\bar M-(m_i+m_j)$ versus \(1/\mu_{ij}\).
		The curve shows the model relation;
		blue circles and red squares denote calculated
		and reference values, respectively.
		The dashed horizontal line marks the saturation limit
		\(-4B_{\rm col}/(3C)+D=176\)~MeV.
		}
		\end{figure}
	
Fig.~\ref{fig:meson_effective_mass_contribution} shows the meson contribution \(-(4/3)F(\mu)+D\), which approaches \(-4B_{\rm col}/(3C)+D=176\)~MeV as \(\mu\to0\), with the rightmost points already close to this plateau. The saturation factor is \(1-e^{-C/\mu}\). Taking \(C=0.439\)~GeV, \(\mu_{n\bar n}\simeq m_n/2\simeq0.117\)~GeV, and \(\mu_{n\bar s}\simeq m_nm_s/(m_n+m_s)\simeq0.153\)~GeV, we obtain \(C/\mu\simeq3.75\) and \(2.87\), corresponding to saturation factors of \(0.976\) and \(0.944\), respectively, so that both light-flavor systems lie close to saturation. In the heavy-flavor region, the attraction instead strengthens with increasing reduced mass.
	
	To quantify the additional attraction relative to light-flavor systems,
	we define
\begin{equation}
	\label{eq:relative_binding_check}
	\begin{aligned}
		\mathcal B_{Q\bar Q'}={}&\bar M_{Q\bar Q'}^{\rm fit}
		-\bar M_{Q\bar n}^{\rm fit}
		-\bar M_{n\bar Q'}^{\rm fit}
		+\bar M_{n\bar n}^{\rm fit}.
	\end{aligned}
\end{equation}
The constituent rest masses and the additive constant \(D\)
	cancel in this combination, so \(\mathcal B_{Q\bar Q'}\) isolates the
	additional pair binding relative to the light-flavor reference.
For $c\bar c$, $b\bar c$, and $b\bar b$, Eq.~\eqref{eq:relative_binding_check} gives
$(-253,-349,-566)$ MeV,
	respectively, close to the corresponding values
	$(-259,-353,-567)$ MeV from Ref.~\cite{Zhang2025Binding},
	listed in Table~\ref{tab:meson_parameters_comparison}
	of the Introduction.
	Both sets show increasing attraction from $c\bar c$
	through $b\bar c$ to $b\bar b$;
	the estimates from Ref.~\cite{KR2018Pwave}
	in the same table follow this ordering as well.

	Light-flavor saturation further connects our mass relation with the effective constituent-mass description. In the common saturation limit, \(F(\mu_{ij})\to F_0=B_{\rm col}/C\). For a color-singlet hadron, the total color generator \(\mathbf T_H=\sum_i\mathbf T_i\) satisfies \(\mathbf T_H|H\rangle=0\). Since every quark or antiquark satisfies \(\mathbf T_i^2=C_F=4/3\), one obtains
	\begin{equation}
		\label{eq:light_saturation_color_sum}
		\sum_{i<j}\langle\mathbf T_i\!\cdot\!\mathbf T_j\rangle_H
		=\frac12\left[\langle\mathbf T_H^2\rangle_H-N_HC_F\right]
		=-\frac{2N_H}{3}.
	\end{equation}
	Thus the leading chromoelectric contribution is proportional to the number of constituents. Defining \(\delta F(\mu_{ij})=F(\mu_{ij})-F_0\) and \(\widetilde m_i=m_i-\frac23F_0\), Eq.~\eqref{eq:spin_average_mass} can be rewritten as
	\begin{equation}
		\label{eq:light_saturation_mass_relation}
		M_{{\rm SI},H}
		=\sum_i\widetilde m_i+D
		+\sum_{i<j}\langle\mathbf T_i\!\cdot\!\mathbf T_j\rangle_H
		\delta F(\mu_{ij}).
	\end{equation}
	The reduced masses are still defined in terms of the calibrated \(m_i\). When the residual pair contribution is small, \(M_{{\rm SI},H}\simeq\sum_i\widetilde m_i+D\), so the leading chromoelectric contribution can be absorbed into the effective constituent masses.
	
In the common light-flavor saturation limit, color algebra reduces the leading chromoelectric contribution to a mass shift shared by all constituents, providing a model basis for absorbing it into the effective constituent masses. This result connects with the treatment in the CMI model, where the averaged spin-independent contribution is incorporated into the effective masses~\cite{Hogaasen2006,Li2025TriplyHeavy,Man2026SinglyHeavy}, and is also compatible with the approach of Karliner and Rosner, in which the light-flavor sector is described by effective masses while heavy--strange and heavy--heavy pairs receive additional binding contributions~\cite{KR2014,KR2018}. The explicitly retained residual term \(\delta F\) captures the pair dependence away from the saturation limit, thereby linking the effective constituent-mass approximation to the flavor-dependent pair corrections within a single mass relation, and connecting the nearly saturated light-flavor contribution to the increasingly attractive behavior in the heavy-flavor region. These results reflect the role of the saturating parametrization in describing the flavor dependence of the chromoelectric contribution, and are consistent with the dynamical treatment of screened potentials in light and heavy meson spectra~\cite{Song2015ScreenedCharm,Wang2021ScreenedLightVectors,Zhang2026RadialLadder}.

	\begin{table*}[!t]
		\caption{\label{tab:stable14_baryon_averages}
			The fit inputs are marked with \(*\), and \(\Delta\bar M=\bar M^{\rm fit}-\bar M^{\rm ref}\). The first entry in parentheses is the \(\pm9\)~MeV predictive scale from the leave-one-out test, and the second is the mass shift from scanning the chromomagnetic couplings (downward and upward).
		}
		\centering
		\renewcommand{\arraystretch}{1.1}
		\begin{ruledtabular}
			\begin{tabular}{lrrrc}
				Flavor & $\bar M^{\rm ref}$
				& $\bar M^{\rm fit}$ & $\Delta\bar M$
				& Reference \\
				\hline
				$nnn^{*}$ & 1134 & $1134(9)(0)$ & $0$ & Exp. \\
				$nns^{*}$ & 1321 & $1318(9)(0)$ & $-3$ & Exp. \\
				$nss$ & 1462 & $1467(9)(0)$ & $+5$ & Exp. \\
				$nnc$ & 2497 & $2487(9)(0)$ & $-10$ & Exp. \\
				$nsc^{*}$ & 2623 & $2616(9)(0)$ & $-8$ & Exp. \\
				$ssc^{*}$ & 2742 & $2742(9)(0)$ & $-1$ & Exp. \\
				$nnb^{*}$ & 5826 & $5831(9)(0)$ & $+5$ & Exp. \\
				$nsb^{*}$ & 5948 & $5949(9)(0)$ & $+2$ & Exp. \\
				$ssb$ & 6075 & $6065(9)(0)$ & $-10$ & Lat. \\
				$ncc$ & 3669 & $3674(9)(5)$ & $+5$ & Exp.+HF \\
				$scc^{*}$ & 3774 & $3779(9)(5)$ & $+5$ & Exp.+HF \\
				$nbb$ & 10166 & $10201(9)(3)$ & $+35$ & Lat. \\
				$sbb$ & 10296 & $10286(9)(3)$ & $-11$ & Lat. \\
				$ncb$ & 6968 & $6960(9)(0)$ & $-8$ & Lat. \\
				$scb$ & 7037 & $7055(9)(0)$ & $+18$ & Lat. \\
				$ccb$ & 8027 & $8021(9)(5)$ & $-6$ & Lat. \\
				$cbb$ & 11218 & $11205(9)(3)$ & $-13$ & Lat. \\
			\end{tabular}
		\end{ruledtabular}
	\end{table*}

	\begin{figure}[htbp]
		\centering
		\includegraphics[width=0.43\textwidth]{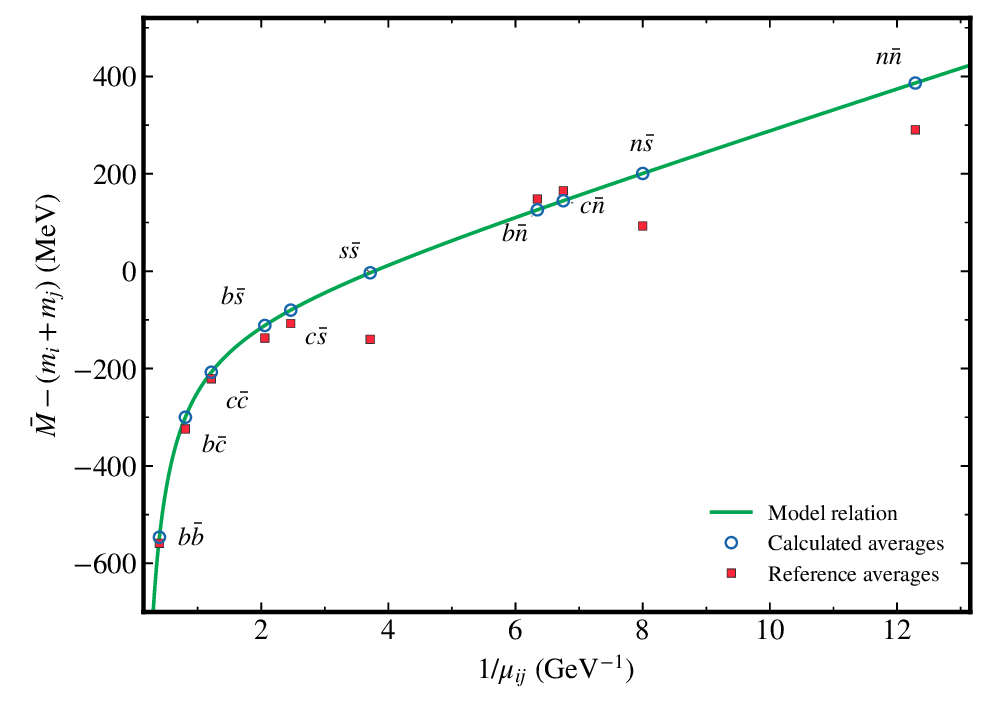}
		\caption{\label{fig:meson_effective_mass_conell}
			Meson contributions $\bar M-(m_i+m_j)$ versus $1/\mu_{ij}$.
			The curve shows the model relation;
			blue circles and red squares denote calculated
			and reference values, respectively.
		}
	\end{figure}
	
As a cross-check, we fix the heavy-quark masses from Table~\ref{tab:stable14_parameters} and fit the Cornell-inspired form
\begin{equation*}
	F_{\rm Cornell}(\mu)
	=A_{\rm Cornell}\mu+\frac{B_{\rm Cornell}}{\mu}
\end{equation*}
to the same 14 inputs with the chromomagnetic couplings held fixed. In this comparison, the hadron-level constant $D$ is replaced by a fitted $D_{\rm Cornell}$. On the \(m_s>m_n\) branch, the joint RMS residual is about 49~MeV, with the meson and baryon sectors contributing about 45 and 52~MeV, respectively; the deviations are concentrated in the light-flavor mesons (see Fig.~\ref{fig:meson_effective_mass_conell}). If all seven parameters are left free subject only to \(m_i>0\) and \(m_s>m_n\), the RMS residual drops to about 17~MeV, but \(m_n\) falls to 0.065~GeV. The leave-one-out predictive RMS is then about 51~MeV, or 29~MeV with the \(b\bar{b}\) input retained; along the eleven- to seventeen-point input sequence, \(m_n\) and \(B_{\rm Cornell}\) vary by up to about \(57\%\) and \(53\%\) relative to their 14-point values. Balancing fit accuracy, parameter values, and stability, we therefore adopt the parameters calibrated with the saturating form in the subsequent calculations.

	\section{Chromomagnetic interaction and conventional hadrons}
	\label{sec:conventional}
	
	\subsection{Couplings and {spin-averaged mass reconstruction}}
	\label{sec:cmi_couplings}
	
	We now specify the fixed chromomagnetic couplings in
	Eq.~\eqref{eq:unified_mass_operator}, whose chromomagnetic
	term is denoted by $H_{\rm CMI}$.

	{
		The couplings for $qq$ and $q\bar q$ pairs are denoted by
		$C_{ij}^{qq}$ and $C_{i\bar j}^{q\bar q}$, respectively.
	}
	The effective couplings $C_{ij}$ have dimensions of energy
	and incorporate the effects of quark masses and pair contact
	probabilities, while the color and spin matrix elements
	remain explicit \cite{DeRujula1975,Weng2018}.
	In potential models, $C_{ij}$ is obtained by averaging
	the radial part of the chromomagnetic interaction over
	the spatial wave function \cite{AnwarBurns2023MassRelations}.
	We calibrate these couplings using hyperfine splittings
	and hold them fixed in the preceding mass fit.
	
	For color-singlet mesons, the spin factors are $-3$ and $+1$
	for pseudoscalar and vector states, respectively;
	the baryon $nn$ coupling is determined from the
	$N$--$\Delta$ splitting.
	The corresponding relations are
	\begin{equation}
		\label{eq:cmi_calibration}
		C_{i\bar j}=\frac{3}{64}(M_V-M_P),\qquad
		C_{nn}=\frac{M_\Delta-M_N}{16}.
	\end{equation}
	{
		Here $M_V$ and $M_P$ are the vector and pseudoscalar
		meson masses for the same flavor pair.
	}
	The meson inputs follow Ref.~\cite{Zhang2025Binding}
	and include the measured $B_c^*-B_c$ splitting
	\cite{ATLAS2026BcStar};
	the $s\bar s$ input uses the unmixed auxiliary state $\eta_s$.
	
	For the remaining baryon $qq$ couplings, we use
	\begin{equation}
		C_{ij}^{qq}\simeq\frac23 C_{i\bar j}^{q\bar q}.
		\label{eq:Cqq-scaling}
	\end{equation}
	The baryon-to-meson effective coupling ratios extracted
	in Ref.~\cite{Weng2018} for the $nn$, $ns$, $nc$, and $sc$
	pairs are approximately $0.600$--$0.700$.
	We adopt the $2/3$ scaling prescription of
	Ref.~\cite{Zhang2025Binding};
	the resulting couplings are listed in Table~\ref{tab:cmi-parameters}.
	
	Pair spatial distributions may differ between tetraquarks
	and baryons. We therefore use the above couplings as a reference
	and examine the dependence of the spectra on the coupling ratio.
	Ref.~\cite{Wu2018FullyHeavy} compared the equal-coupling
	choice $C_{QQ}=C_{Q\bar Q}$ with a prescription based on
	the light-flavor coupling ratio for fully heavy tetraquarks.
	For a tetraquark with a single flavor, equal couplings
	also follow from the approximation that the spatial wave function
	is symmetric under exchanges of quark and antiquark coordinates
	\cite{AnwarBurns2023MassRelations}.

		For the central spectra, we use the values in
		Table~\ref{tab:cmi-parameters}.
		We examine the coupling dependence through
		$C_{ij}^{qq}=rC_{i\bar j}^{q\bar q}$ for
		$ij=nc,sc,nb,sb,cc,cb,bb$, using a common ratio
		$r\in[1/3,1]$ for these pairs.
		This range covers the empirical value $r\simeq2/3$
		\cite{Weng2018,Zhang2025Binding}
		and the equal-coupling choice $r=1$
		\cite{Wu2018FullyHeavy,AnwarBurns2023MassRelations},
		and extends to weaker $qq$ contact strengths.
		Antiquark pairs use the corresponding $qq$ coefficients,
		while all $q\bar q$ coefficients and light-flavor
		$qq$ coefficients remain fixed.
		The mass parameters are held fixed throughout the scan,
		and the complete mass matrices are diagonalized again at each $r$.

	\begin{table}[!t]
		\centering
		\caption{\label{tab:cmi-parameters}
			Fixed chromomagnetic couplings used in the spectra (MeV).
			Color and spin factors are included separately in the operator;
			$n=u,d$.
		}
		\renewcommand{\arraystretch}{1.0}
		\begin{ruledtabular}
			\begin{tabular}{crr}
				Flavor pair $ij$ & $C_{i\bar j}$ ($q\bar q$)
				& $C_{ij}$ ($qq$) \\
				\hline
				$nn$ & 29.861 & 18.313 \\
				$ns$ & 18.662 & 12.441 \\
				$ss$ & 15.588 & 10.394 \\
				$nc$ & 6.623 & 4.416 \\
				$sc$ & 6.741 & 4.495 \\
				$cc$ & 5.288 & 3.525 \\
				$nb$ & 2.118 & 1.412 \\
				$sb$ & 2.274 & 1.516 \\
				$cb$ & 3.023 & 2.016 \\
				$bb$ & 2.892 & 1.928 \\
			\end{tabular}
		\end{ruledtabular}
	\end{table}
	
	For conventional hadrons, the spin-independent contribution
	is common to the spin partners, giving
	\begin{equation}
		\label{eq:total_mass}
		M_{H,k}^{\rm cal}
		=M_{{\rm SI},H}+E_{H,k}.
	\end{equation}
Here $E_{H,k}$ is an eigenvalue of $H_{\rm CMI}$, and
	$M_{{\rm SI},H}$ is given by Eq.~\eqref{eq:spin_average_mass}.

	{
		For the selected multiplet, we decompose the chromomagnetic
		contribution into the average $\bar E_H$ defined in
		Eq.~\eqref{eq:cmi_average} and a relative shift,
		\begin{equation}
			\label{eq:cmi_centering}
			\delta E_{H,k}=E_{H,k}-\bar E_H.
		\end{equation}
		The individual-state masses can equivalently be written as
		$M_{H,k}^{\rm cal}=\bar M_H^{\rm fit}+\delta E_{H,k}$\revblue{.}
	}

	\subsection{Baryon spectra}
	
	Using the calibrated parameters and the couplings in
	Table~\ref{tab:cmi-parameters}, we calculate the ground-state
	baryon spectrum from Eq.~\eqref{eq:unified_mass_operator},
	which takes the form of Eq.~\eqref{eq:total_mass}
	for conventional hadrons.
	Since the meson couplings are calibrated to pseudoscalar--vector
	splittings, the individual meson mass residuals primarily reflect
	the spin-averaged mass differences in
	Table~\ref{tab:stable14_meson_averages}.
	We therefore focus on the baryon results to further examine
	the $qq$ couplings and their flavor dependence.
	
	In the tables, $M(9)({}^{+u}_{-l})$ denotes the central mass,
	the empirical $\pm9$ MeV error estimated from the leave-one-out test,
	and the downward and upward coupling-scan shifts $l,u\geq0$,
	respectively.
	Together, these contributions span $[M-9-l,M+9+u]$.
	A single number in the second parentheses indicates a symmetric shift;
	zero denotes a change below the tabulated precision.
	This convention also applies to the spin-averaged masses
	and the subsequent tetraquark results.
	During the scan, the spin-independent masses are held fixed,
	and the individual-state masses are recalculated using
	Eq.~\eqref{eq:total_mass}.
	The corresponding average CMI contributions also vary with
	$r=C_{ij}^{qq}/C_{i\bar j}^{q\bar q}$,
	the ratio of the $qq$ coupling to the meson $q\bar q$ coupling
	for the same flavor pair.

	For the $\Sigma_Q^{(*)}$, $\Omega_Q^{(*)}$, and $qQQ$ spin doublets,
	we denote the spin-one quark pair by $q_1q_1$ and the remaining
	quark by $q_2$, where $Q=c,b$ and $q=n,s$.
	Each quark pair has a color factor of $-2/3$, giving
	\begin{equation}
		\label{eq:cmi_baryon_pair}
		\begin{aligned}
			E_{1/2}&=\frac83(C_{q_1q_1}-4C_{q_1q_2}),\\
			E_{3/2}&=\frac83(C_{q_1q_1}+2C_{q_1q_2}).
		\end{aligned}
	\end{equation}
	For $nnQ$, we set $q_1=n$ and $q_2=Q$;
	for $qQQ$, we set $q_1=Q$ and $q_2=q$.
	The spin-weighted CMI average of the two states is
	$8C_{q_1q_1}/3$, with relative shifts
	$\delta E_{1/2}=-32C_{q_1q_2}/3$
	and $\delta E_{3/2}=16C_{q_1q_2}/3$.
	Subtracting the two masses gives
	$M_{3/2}-M_{1/2}=16C_{q_1q_2}$.
	The hyperfine splitting thus directly tests the $q_1q_2$ coupling
	after the common mass contribution cancels.
	
	For $nnQ$, these expressions apply to $\Sigma_Q$ and $\Sigma_Q^*$.
	The light-quark pair in $\Lambda_Q$ has spin zero,
	giving $E_{\Lambda_Q}=-8C_{nn}$.
	Relative to the spin-weighted $\Sigma_Q,\Sigma_Q^*$ average,
	its shift is $\delta E_{\Lambda_Q}=-32C_{nn}/3$.
	
	For baryons with three distinct flavors, we label the quarks by $1,2,3$.
	In the $S_{12}=0,1$ basis, the spin-$1/2$ CMI matrix is
	\cite{Weng2018}
	\begin{equation}
		\label{eq:cmi_baryon_mixing}
		\mathcal H_{1/2}=\frac83
		\begin{pmatrix}
			-3C_{12}&\sqrt3(C_{23}-C_{13})\\
			\sqrt3(C_{23}-C_{13})&C_{12}-2(C_{13}+C_{23})
		\end{pmatrix},
	\end{equation}
	where $C_{ij}$ is the coupling between quarks $i$ and $j$.
	The spin-$3/2$ contribution is
	$E_{3/2}=8(C_{12}+C_{13}+C_{23})/3$.
	After diagonalization, the average contribution and relative shifts
	are calculated using Eqs.~\eqref{eq:cmi_average}
	and \eqref{eq:cmi_centering}, respectively.
	
	For $nsQ$, we take $(1,2,3)=(n,s,Q)$
	and use the spin-weighted average of the higher spin-$1/2$ state
	$\Xi'_Q$ and the spin-$3/2$ state $\Xi_Q^*$ as the mass baseline.
	The lower $\Xi_Q$ state is also expressed relative to this baseline.
	For $ncb$ and $scb$, including all three levels with weights $1:1:2$
	gives a vanishing CMI average.
	The labels $L,H$ denote the lower and higher spin-$1/2$ eigenstates,
	respectively.
	
	For $sss$, $ccc$, and $bbb$, the ground state
	has $J^P=3/2^+$, with mass
	\begin{equation}
		\label{eq:cmi_identical_baryon}
		M_{fff}^{\rm cal}
		=M_{{\rm SI},fff}+8C_{ff},
		\qquad f=s,c,b,
	\end{equation}
	where $M_{{\rm SI},fff}$ is obtained by extrapolating
	Eq.~\eqref{eq:spin_average_mass}.
	These states are marked by daggers in the tables.
	
	The results are listed in
	Table~\ref{tab:stable14_baryon_spectrum_exp}
	and Table~\ref{tab:stable14_heavy_baryon_spectrum}.
	Experimental masses are taken from Ref.~\cite{PDG2026},
	with $M_\Delta=1232$ MeV and the negatively charged state
	used for $\Xi'_b$.
	For $\Omega_{cc}^{+}$, we use the measurement in
	Ref.~\cite{LHCb2026OmegaCC}.
	The lattice reference values are taken from
	Ref.~\cite{Brown2014Baryons}.
	
	\begin{table*}[htbp]
		\caption{\label{tab:stable14_baryon_spectrum_exp}
			Masses are in MeV.
			$\delta E=M^{\rm cal}-M_{\rm base}$;
			$\delta M=M^{\rm cal}-M^{\rm ref}$.
			Daggers mark states evaluated using Eq.~\eqref{eq:cmi_identical_baryon}.
			Exp and Lat denote experimental and lattice references.
			The two sets of parentheses give the $\pm9$ MeV empirical
			prediction scale and coupling-scan shifts.
		}
		\centering
		\small
		\setlength{\tabcolsep}{3pt}
		\renewcommand{\arraystretch}{0.8}
		\begin{ruledtabular}
			\begin{tabular}{lcrrrrr}
				State & $J^P$ & $M_{\rm base}$ & $\delta E$
				& $M^{\rm cal}$ & $M^{\rm ref}$
				& $\delta M$ \\
				\hline
				$N$ & $\frac12^+$
				& 1134 & $-194$ & $939(9)(0)$
				& $939^{\rm Exp}$ & $0$ \\
				$\Delta$ & $\frac32^+$
				& 1134 & $+97$ & $1231(9)(0)$
				& $1232^{\rm Exp}$ & $-1$ \\
				$\Lambda$ & $\frac12^+$
				& 1318 & $-194$ & $1123(9)(0)$
				& $1116^{\rm Exp}$ & $+7$ \\
				$\Sigma$ & $\frac12^+$
				& 1318 & $-133$ & $1185(9)(0)$
				& $1193^{\rm Exp}$ & $-8$ \\
				$\Sigma^*$ & $\frac32^+$
				& 1318 & $+66$ & $1384(9)(0)$
				& $1385^{\rm Exp}$ & $-1$ \\
				$\Xi$ & $\frac12^+$
				& 1467 & $-133$ & $1334(9)(0)$
				& $1318^{\rm Exp}$ & $+16$ \\
				$\Xi^*$ & $\frac32^+$
				& 1467 & $+66$ & $1533(9)(0)$
				& $1533^{\rm Exp}$ & $0$ \\
				$\Omega^\dagger$ & $\frac32^+$
				& 1597 & $+83$ & $1680(9)(0)$
				& $1672^{\rm Exp}$ & $+7$ \\
				\hline
				$\Lambda_c$ & $\frac12^+$
				& 2487 & $-194$ & $2292(9)(0)$
				& $2286^{\rm Exp}$ & $+6$ \\
				$\Sigma_c$ & $\frac12^+$
				& 2487 & $-47$ & $2439(9)(24)$
				& $2453^{\rm Exp}$ & $-14$ \\
				$\Sigma_c^*$ & $\frac32^+$
				& 2487 & $+24$ & $2510(9)(12)$
				& $2518^{\rm Exp}$ & $-8$ \\
				$\Xi_c$ & $\frac12^+$
				& 2616 & $-133$ & $2483(9)(0)$
				& $2469^{\rm Exp}$ & $+14$ \\
				$\Xi_c'$ & $\frac12^+$
				& 2616 & $-48$ & $2568(9)(24)$
				& $2579^{\rm Exp}$ & $-10$ \\
				$\Xi_c^*$ & $\frac32^+$
				& 2616 & $+24$ & $2640(9)(12)$
				& $2646^{\rm Exp}$ & $-6$ \\
				$\Omega_c$ & $\frac12^+$
				& 2742 & $-48$ & $2694(9)(24)$
				& $2695^{\rm Exp}$ & $-2$ \\
				$\Omega_c^*$ & $\frac32^+$
				& 2742 & $+24$ & $2766(9)(12)$
				& $2766^{\rm Exp}$ & $0$ \\
				\hline
				$\Lambda_b$ & $\frac12^+$
				& 5831 & $-194$ & $5636(9)(0)$
				& $5620^{\rm Exp}$ & $+17$ \\
				$\Sigma_b$ & $\frac12^+$
				& 5831 & $-15$ & $5816(9)(8)$
				& $5813^{\rm Exp}$ & $+2$ \\
				$\Sigma_b^*$ & $\frac32^+$
				& 5831 & $+8$ & $5838(9)(4)$
				& $5833^{\rm Exp}$ & $+6$ \\
				$\Xi_b$ & $\frac12^+$
				& 5949 & $-133$ & $5817(9)(0)$
				& $5794^{\rm Exp}$ & $+22$ \\
				$\Xi_b'$ & $\frac12^+$
				& 5949 & $-16$ & $5934(9)(8)$
				& $5935^{\rm Exp}$ & $-1$ \\
				$\Xi_b^*$ & $\frac32^+$
				& 5949 & $+8$ & $5957(9)(4)$
				& $5954^{\rm Exp}$ & $+3$ \\
				$\Omega_b$ & $\frac12^+$
				& 6065 & $-16$ & $6049(9)(8)$
				& $6046^{\rm Exp}$ & $+3$ \\
				$\Omega_b^*$ & $\frac32^+$
				& 6065 & $+8$ & $6073(9)(4)$
				& $6085(47)(20)^{\rm Lat}$ & $-12$ \\
			\end{tabular}
		\end{ruledtabular}
	\end{table*}
	
	\begin{table*}[htpb]
		\caption{\label{tab:stable14_heavy_baryon_spectrum}
			Doubly and triply heavy baryon masses (MeV).
			Notation follows Table~\ref{tab:stable14_baryon_spectrum_exp}.
			$L,H$ label the lower and higher mixed spin-$1/2$ levels.
			Lattice uncertainties are statistical and systematic, respectively.
		}
		\centering
		\small
		\setlength{\tabcolsep}{3pt}
		\renewcommand{\arraystretch}{0,9}
		\begin{ruledtabular}
			\begin{tabular}{lcrrrrr}
				State & $J^P$ & $M_{\rm base}$ & $\delta E$
				& $M^{\rm cal}$ & $M^{\rm ref}$
				& $\delta M$ \\
				\hline
				$\Xi_{cc}$ & $\frac12^+$
				& 3674 & $-47$ & $3627(9)(19)$
				& $3622^{\rm Exp}$ & $+5$ \\
				$\Xi_{cc}^*$ & $\frac32^+$
				& 3674 & $+24$ & $3697(9)(16)$
				& $3692(28)(21)^{\rm Lat}$ & $+5$ \\
				$\Omega_{cc}$ & $\frac12^+$
				& 3779 & $-48$ & $3731(9)(19)$
				& $3726^{\rm Exp}$ & $+5$ \\
				$\Omega_{cc}^*$ & $\frac32^+$
				& 3779 & $+24$ & $3803(9)(17)$
				& $3822(20)(22)^{\rm Lat}$ & $-19$ \\
				$\Xi_{bb}$ & $\frac12^+$
				& 10201 & $-15$ & $10186(9)(5)$
				& $10143(30)(23)^{\rm Lat}$ & $+43$ \\
				$\Xi_{bb}^*$ & $\frac32^+$
				& 10201 & $+8$ & $10209(9)(6)$
				& $10178(30)(24)^{\rm Lat}$ & $+31$ \\
				$\Omega_{bb}$ & $\frac12^+$
				& 10286 & $-16$ & $10270(9)(6)$
				& $10273(27)(20)^{\rm Lat}$ & $-3$ \\
				$\Omega_{bb}^*$ & $\frac32^+$
				& 10286 & $+8$ & $10294(9)(7)$
				& $10308(27)(21)^{\rm Lat}$ & $-14$ \\
				\hline
				$\Xi_{cb,L}$ & $\frac12^+$
				& 6960 & $-36$ & $6925(9)(18)$
				& $6943(33)(28)^{\rm Lat}$ & $-18$ \\
				$\Xi_{cb,H}$ & $\frac12^+$
				& 6960 & $-6$ & $6954(9)(3)$
				& $6959(36)(28)^{\rm Lat}$ & $-5$ \\
				$\Xi_{cb}^*$ & $\frac32^+$
				& 6960 & $+21$ & $6981(9)(10)$
				& $6985(36)(28)^{\rm Lat}$ & $-4$ \\
				$\Omega_{cb,L}$ & $\frac12^+$
				& 7055 & $-36$ & $7019(9)(18)$
				& $6998(27)(20)^{\rm Lat}$ & $+21$ \\
				$\Omega_{cb,H}$ & $\frac12^+$
				& 7055 & $-7$ & $7049(9)(3)$
				& $7032(28)(20)^{\rm Lat}$ & $+17$ \\
				$\Omega_{cb}^*$ & $\frac32^+$
				& 7055 & $+21$ & $7077(9)(11)$
				& $7059(28)(21)^{\rm Lat}$ & $+18$ \\
				\hline
				$\Omega_{ccb}$ & $\frac12^+$
				& 8021 & $-22$ & $8000(9)(6)$
				& $8007(9)(20)^{\rm Lat}$ & $-7$ \\
				$\Omega_{ccb}^*$ & $\frac32^+$
				& 8021 & $+11$ & $8032(9)(10)$
				& $8037(9)(20)^{\rm Lat}$ & $-5$ \\
				$\Omega_{cbb}$ & $\frac12^+$
				& 11205 & $-22$ & $11183(9)(8)$
				& $11195(8)(20)^{\rm Lat}$ & $-12$ \\
				$\Omega_{cbb}^*$ & $\frac32^+$
				& 11205 & $+11$ & $11215(9)(8)$
				& $11229(8)(20)^{\rm Lat}$ & $-14$ \\
				$\Omega_{ccc}^\dagger$ & $\frac32^+$
				& 4764 & $+28$ & $4793(9)(14)$
				& $4796(8)(18)^{\rm Lat}$ & $-3$ \\
				$\Omega_{bbb}^\dagger$ & $\frac32^+$
				& 14327 & $+15$ & $14342(9)(8)$
				& $14366(9)(20)^{\rm Lat}$ & $-24$ \\
			\end{tabular}
		\end{ruledtabular}
	\end{table*}
	
	For light and singly heavy baryons, the calculated masses
	differ from experiment by at most 22 MeV.
	The $\Xi_{cc}$, which is not included in the 14-point fit,
	has a predicted mass of 3627 MeV, only 5 MeV above experiment.
	Its spin partner $\Xi_{cc}^*$ also differs from the lattice result
	by 5 MeV \cite{Brown2014Baryons}.
	Our $\Omega_{cc}^*$ prediction of 3803 MeV is close to
	the lattice value of $3788(13)(12)$ MeV
	\cite{MathurPadmanath2019OmegaCC}.
	
	The doubly bottom baryon masses lie in a range similar to
	the model predictions in Refs.~\cite{KR2014,Ebert2002DoublyHeavy}.
	The $\Omega_{bb}$ and $\Omega_{bb}^*$ masses differ from
	the lattice results by $-3$ and $-14$ MeV, respectively.
	For the $ncb$ and $scb$ systems with spin mixing,
	the deviations of the individual levels do not exceed
	18 and 21 MeV, respectively, and remain within
	the corresponding lattice uncertainties \cite{Brown2014Baryons}.
	These comparisons extend the tests of the unified mass relation
	to heavy-flavor combinations outside the calibration set.
	
	For triply heavy baryons,
	Eq.~\eqref{eq:cmi_identical_baryon} gives
	$M_{\Omega_{ccc}}=4793$ MeV,
	matching the central value of the continuum-extrapolated lattice result
	$4793(5)({}^{+11}_{-8})$ MeV \cite{Dhindsa2025OmegaCCC}.
	Our prediction $M_{\Omega_{bbb}}=14342$ MeV is
	24--29 MeV below the two lattice determinations
	\cite{Brown2014Baryons,Meinel2010OmegaBBB}.
	The mass suppression of heavy--heavy chromomagnetic couplings
	makes their contributions small relative to the spin-independent masses.
	Over the coupling range considered,
	the $\Omega_{ccc}$ and $\Omega_{bbb}$ masses shift
	by about 14 and 8 MeV, respectively,
	while both members of the $\Xi_{bb}$ doublet shift
	by at most about 6 MeV.
	
	These baryon results show that a single calibrated parameter set
	can predict masses across different flavor combinations.
	In the original extended CMI formulation,
	constituent masses and chromoelectric contributions are combined
	into effective pair masses, with separate parameters for
	different flavor combinations in the $q\bar q$ and $qq$ sectors
	\cite{Weng2018,Weng2021}.
	Their values are extracted from known hadron masses
	or estimated using supplementary relations when experimental
	constraints are insufficient.
	Flavor dependence is thus encoded in the values
	of the individual pair parameters.
	
	Our framework links different flavor pairs through the reduced mass,
	while explicit color factors account for the color structure
	of the $q\bar q$ and $qq$ pairs.
	With the chromomagnetic couplings held fixed,
	a joint fit to meson and baryon masses determines eight
	spin-independent parameters that simultaneously specify
	the chromoelectric contributions of all flavor pairs,
	without introducing additional pair-specific parameters.
	These cross-flavor correlations allow masses from different
	hadron sectors to constrain the model jointly
	and extend the calibration constraints to mass predictions
	for unfitted baryons and multiquark states.
We next apply the same operator to tetraquarks and examine
	their spectra and dependence on the chromomagnetic couplings.
	
	\section{Mass spectra of tetraquark states}
	\label{sec:tetraquarks}
	
	\subsection{Calculation method and comparison conventions}
	\label{sec:tetra_method}

In this section we extend the above mass relation to fully heavy tetraquarks, treating the chromoelectric and chromomagnetic contributions in color--spin bases that respect exchange symmetry. Using the parameters in Table~\ref{tab:stable14_parameters} and Table~\ref{tab:cmi-parameters}, we apply Eq.~\eqref{eq:unified_mass_operator} to \(q_1q_2\bar q_1\bar q_2\) with \((q_1,q_2)=(c,c),(b,b),(b,c)\). In each \(J^{PC}\) channel, the mass matrix is
\begin{equation}
	\label{eq:tetra_full_matrix}
	\mathcal M^{J^{PC}}
	=[\mathcal M_{\rm SI}\otimes\mathbf I_s]_{J^{PC}}
	+\mathcal H_{\rm CMI}^{J^{PC}},
\end{equation}
where \(\mathcal M_{\rm SI}\) is the color-space matrix in Eq.~\eqref{eq:app_si_matrix}, \(\mathbf I_s\) is the identity in spin space, and \(\mathcal H_{\rm CMI}^{J^{PC}}\) is the chromomagnetic matrix in the same basis. The brackets denote projection onto the basis allowed by exchange symmetry; the bases and matrices are given in Appendix~\ref{app:basis} and Appendix~\ref{app:matrices}. Diagonalization yields \(M_{\rm mix}=M_k^{\rm cal}\) and the eigenvectors. Component expectation values are computed from these eigenvectors, and spin averages include all Pauli-allowed levels with weights \(2J+1\).

Restricting the color space to \(|\bar33\rangle\) or \(|6\bar6\rangle\) gives \(M_{\bar33}\) or \(M_{6\bar6}\), retaining spin mixing within each block. For each calculation, \(k\) orders the masses within each \(J^{PC}\) channel. The probabilities \(P_{\bar33}\) and \(P_{6\bar6}=1-P_{\bar33}\) are summed over the spin components in each color sector and are quoted in percent.

Comparisons use \(\delta M=M^{\rm cal}-M^{\rm ref}\) and \(\Delta_{M_1M_2}=M_{\rm mix}-M_{M_1}^{\rm ref}-M_{M_2}^{\rm ref}\), with the meson reference masses listed in the spectrum tables.

	\subsection{Fully charmed and fully bottom tetraquarks}
	\label{sec:fullyheavy14}
	
	For $QQ\bar Q\bar Q$ systems with $Q=c,b$, every quark pair has the same reduced mass $m_Q/2$, and the color factors of all pairs sum to $-8/3$. The spin-independent part of Eq.~\eqref{eq:unified_mass_operator} therefore yields the same mass for both color configurations
	\begin{equation}
		\label{eq:fullyheavy_baseline14}
		\begin{aligned}
			M_{{\rm SI},QQ\bar Q\bar Q}
			&=4m_Q-\frac83F(m_Q/2)+D\\
			&=2\bar M_{Q\bar Q}^{\rm fit}-D,
		\end{aligned}
	\end{equation}
	where $\bar M_{Q\bar Q}^{\rm fit}$ is the fitted quarkonium
	spin average, whose mean chromomagnetic contribution vanishes.
	With this average fixed, the absolute mass scale tests
	the continuation of the jointly calibrated constant $D$
	to tetraquarks.
	Using the calibrated parameters, we obtain
	$M_{{\rm SI},cc\bar c\bar c}=6458$ MeV
	and $M_{{\rm SI},bb\bar b\bar b}=19208$ MeV.
	
	Pauli symmetry allows two $0^{++}$ levels and one each
	with $1^{+-}$ and $2^{++}$.
	The two scalar states arise from mixing
	$|\bar33;11;0\rangle$ and $|6\bar6;00;0\rangle$;
	the corresponding chromomagnetic matrix is given in
	Eq.~\eqref{eq:app_fullyheavy_cmi}.
	According to Eq.~\eqref{eq:app_equal_centroid},
	the {spin-averaged masses} of the fully charmed
	and fully bottom systems are
	${6478}$ and ${19219}$ MeV, respectively.
	
	Table~\ref{tab:fullyheavy14_spectrum_comparison}
	summarizes the calculated spectra, color probabilities,
	and threshold offsets, together with comparisons to other
	compact tetraquark models and lattice QCD.
	These include the Set I CMI model \cite{Wu2018FullyHeavy},
	its chromoelectric extension \cite{Weng2021},
	the MIT bag model \cite{Yan2023Binding},
	the effective diquark spin-interaction model \cite{Xia2026FullyHeavy},
	the relativistic diquark model \cite{Faustov2022FullyHeavy},
	and the nonrelativistic potential model \cite{Liu2024FullyHeavy}.
	Panel (b) also includes the recent quark-model analysis of
	Ref.~\cite{AnwarBurns2026ScalarPartner} and the lattice QCD results
	of Ref.~\cite{Li2026ScalarTensorLattice}.
	
	\begin{table*}[!t]
		\caption[Fully charmed and fully bottom tetraquark spectra.]{\label{tab:fullyheavy14_spectrum_comparison}
			Fully charmed and fully bottom tetraquarks: (a) spectra, color-component probabilities, and threshold shifts; (b) comparison with model and lattice QCD results. Masses, shifts, and \(w_{ki}\) are in MeV, and probabilities are given in percent. Parentheses in the headings denote the thresholds, and those following the shifts denote the \(S\)-wave weights from Eq.~\eqref{eq:swave_rearrangement_width} at the central parameters. A dash indicates an unavailable result or a forbidden channel, and the superscript \(D\) denotes a \(D\)-wave channel.
		}
		\centering
		\footnotesize
		\setlength{\tabcolsep}{1.8pt}
		\renewcommand{\arraystretch}{1.13}
		\begin{ruledtabular}
			\begin{tabular}{lccrrrrrrr}
				\multicolumn{10}{c}{(a) Present results and threshold offsets}\\
				Flavor & $J^{PC}$ & $k$ & $M_{\bar33}$ & $M_{6\bar6}$
				& $M_{\rm mix}$ & $P_{\bar33}$
				& \shortstack{$\Delta_{\eta_c\eta_c}$\\$(5968)$}
				& \shortstack{$\Delta_{\eta_cJ/\psi}$\\$(6081)$}
				& \shortstack{$\Delta_{J/\psi J/\psi}$\\$(6194)$}\\
				\hline
				\multirow{4}{*}{$cc\bar c\bar c$} & \multirow{2}{*}{$0^{++}$} & 1 & \multirow{2}{*}{{${6420}{(9)(9)}$}} & \multirow{2}{*}{{${6486}{(9)(14)}$}} & {${6344}{(9)(11)}$} & 65.130 & ${{+376}}(446.9)$ & --- & ${{+151}}(23.4)$\\
				&  & 2 &  &  & {${6562}{(9)(12)}$} & 34.870 & ${{+594}}(1.6)$ & --- & ${{+368}}(595.2)$\\
				& $1^{+-}$ & --- & {${6448}{(9)(9)}$} & --- & {${6448}{(9)(9)}$} & 100.000 & --- & ${{+367}}(357.3)$ & ---\\
				& $2^{++}$ & --- & {${6505}{(9)(9)}$} & --- & {${6505}{(9)(9)}$} & 100.000 & ${{+537}}^{D}$ & --- & ${{+311}}(331.3)$\\
				\hline
				Flavor & $J^{PC}$ & $k$ & $M_{\bar33}$ & $M_{6\bar6}$
				& $M_{\rm mix}$ & $P_{\bar33}$
				& \shortstack{$\Delta_{\eta_b\eta_b}$\\$(18797)$}
				& \shortstack{$\Delta_{\eta_b\Upsilon}$\\$(18859)$}
				& \shortstack{$\Delta_{\Upsilon\Upsilon}$\\$(18921)$}\\
				\hline
				\multirow{4}{*}{$bb\bar b\bar b$} & \multirow{2}{*}{$0^{++}$} & 1 & \multirow{2}{*}{{${19187}{(9)(5)}$}} & \multirow{2}{*}{{${19223}{(9)(8)}$}} & {${19146}{(9)(6)}$} & 65.130 & ${{+348}}(755.6)$ & --- & ${{+225}}(49.9)$\\
				&  & 2 &  &  & {${19265}{(9)(7)}$} & 34.870 & ${{+467}}(2.5)$ & --- & ${{+344}}(995.7)$\\
				& $1^{+-}$ & --- & {${19203}{(9)(5)}$} & --- & {${19203}{(9)(5)}$} & 100.000 & --- & ${{+344}}(603.0)$ & ---\\
				& $2^{++}$ & --- & {${19234}{(9)(5)}$} & --- & {${19234}{(9)(5)}$} & 100.000 & ${{+436}}^{D}$ & --- & ${{+313}}(576.1)$\\
			\end{tabular}
		\end{ruledtabular}
		\medskip
		\begingroup
		\scriptsize
		\setlength{\tabcolsep}{2pt}
		\begin{ruledtabular}
			\begin{tabular}{lccrrrrrrrrr}
				\multicolumn{12}{c}{(b) Model and lattice QCD mass comparison}\\
				Flavor & $J^{PC}$ & $k$ & Present
				& \shortstack{CMI\\\cite{Wu2018FullyHeavy}}
				& \shortstack{ECMI\\\cite{Weng2021}}
				& \shortstack{Bag\\\cite{Yan2023Binding}}
				& \shortstack{Diq.\\\cite{Xia2026FullyHeavy}}
				& \shortstack{RQM\\\cite{Faustov2022FullyHeavy}}
				& \shortstack{NRPQM\\\cite{Liu2024FullyHeavy}}
				& \shortstack{Ref.\\\cite{AnwarBurns2026ScalarPartner}}
				& \shortstack{Lattice\\\cite{Li2026ScalarTensorLattice}}\\
				\hline
				\multirow{4}{*}{$cc\bar c\bar c$} & \multirow{2}{*}{$0^{++}$} & 1 & {${6344}{(9)(11)}$} & 6035 & 6045 & 6469 & 6098 & \multirow{2}{*}{6190} & 6455 & $6443(19)$ & ---\\
				&  & 2 & {${6562}{(9)(12)}$} & 6254 & 6271 & 6572 & 6314 &  & 6550 & $6650(19)$ & ---\\
				& $1^{+-}$ & --- & {${6448}{(9)(9)}$} & 6137 & 6231 & 6519 & 6204 & 6271 & 6500 & $6540(16)$ & ---\\
				& $2^{++}$ & --- & {${6505}{(9)(9)}$} & 6194 & 6287 & 6545 & 6261 & 6367 & 6524 & $6593(15)^{\dagger}$ & \shortstack{$6538(13)$\\$6543(10)$}\\
				\hline
				\multirow{4}{*}{$bb\bar b\bar b$} & \multirow{2}{*}{$0^{++}$} & 1 & {${19146}{(9)(6)}$} & 18834 & 18836 & 19685 & 18949 & \multirow{2}{*}{19315} & 19306 & --- & ---\\
				&  & 2 & {${19265}{(9)(7)}$} & 18954 & 18981 & 19717 & 19067 &  & 19355 & --- & ---\\
				& $1^{+-}$ & --- & {${19203}{(9)(5)}$} & 18890 & 18969 & 19700 & 19007 & 19320 & 19329 & --- & ---\\
				& $2^{++}$ & --- & {${19234}{(9)(5)}$} & 18921 & 19000 & 19708 & 19038 & 19331 & 19341 & --- & ---\\
			\end{tabular}
		\end{ruledtabular}
		\endgroup

	\end{table*}
	
	Relative to the calculation restricted to
	$\bar3_c\otimes3_c$, configuration mixing lowers the
	lightest scalar masses by 76 MeV for charm and 41 MeV for bottom.
	Both states retain an antitriplet--triplet probability of about $65\%$.

	The recent quark-model analysis of
	Ref.~\cite{AnwarBurns2026ScalarPartner} uses the CMS tensor mass
	as input to calculate its partner states.
	For $R=C_{c\bar c}/C_{cc}=1$, it gives scalar--tensor splittings of
	$M_{2^{++}}-M_{0^{++}_1}=150$ MeV and
	$M_{0^{++}_2}-M_{2^{++}}=57$ MeV,
	in close agreement with our corresponding results of 161 and 57 MeV.
	Both calculations also yield the same level ordering,
	$M_{0^{++}_1}<M_{1^{+-}}<M_{2^{++}}<M_{0^{++}_2}$.
	The agreement in splittings and level ordering supports
	our description of the color--spin multiplet structure.
	
	Our tensor mass is 6505 MeV.
	Including the empirical $\pm9$ MeV prediction scale from
	the leave-one-out test and the chromomagnetic coupling scan
	gives a mass range of 6486--6523 MeV.
	Recent lattice QCD calculations yield
	$6538(13)$ and $6543(10)$ MeV in the same $2^{++}$ channel
	\cite{Li2026ScalarTensorLattice}.
	These lattice central values differ from ours by approximately
	$0.5\%$--$0.6\%$, indicating similar tensor mass scales
	around 6.5 GeV in the fully charmed system.
	
	For the fully charmed system, the upper scalar and tensor
	masses are ${6562}$ and ${6505}$ MeV,
	respectively, in the mass region near $X(6600)$.
	The CMS interference fit to the Run~2+3 data gives
	$M_{X(6600)}=6593^{+15}_{-14}\pm25$ MeV,
	whereas the noninterference fit gives $6540\pm8$ MeV,
	with the latter uncertainty being statistical only
	\cite{CMS2026AllCharmFamily}.
	Relative to these two central values, the upper scalar
	has mass differences of $-31$ and $+22$ MeV,
	and the tensor has differences of $-88$ and $-35$ MeV.
	The upper scalar is therefore closer in mass to both results.
	The difference between the two fitted central values also
	illustrates the effect of interference on the extracted mass.
	We use the experimentally preferred interference fit
	as the primary comparison.
	
	Regarding quantum numbers, the CMS angular analysis favors
	$J^{PC}=2^{++}$ under the common-quantum-number hypothesis
	and the tested amplitude models \cite{CMS2025SpinParity},
	consistent with the quantum numbers of our tensor state.
	Thus, the model predicts two levels with different spin
	correlations in this mass region, providing a spectroscopic
	basis for further studies of scalar and tensor contributions.
	
	In the scan that fixes $C_{c\bar c}$ and varies $C_{cc}$
	through $r\in[1/3,1]$, the upper scalar and tensor masses
	range over 6549--6574 and 6495--6514 MeV, respectively.
	Their maximum shifts from the central results are about
	13 and 10 MeV.
	The level ordering remains unchanged, and the lower scalar
	remains dominated by the antitriplet--triplet color component.
	Leave-one-out refits give a tensor mass range of
	6492--6516 MeV, likewise indicating limited sensitivity
	to changes in the calibration inputs.
	The chromoelectric term in Eq.~\eqref{eq:fullyheavy_baseline14}
	makes a non-negligible contribution to the overall mass scale,
	while the chromomagnetic term produces the spin splittings
	and configuration mixing.
	
	Following the $S$-wave rearrangement-decay analyses in
	Refs.~\cite{Weng2021,Liu2025CompactPentaquarks}, we use the central
	couplings and mass eigenstates to estimate partial-width ratios
	for the fully charmed and fully bottom systems. The partial width is
	\begin{equation}
		\label{eq:swave_rearrangement_width}
		\Gamma_{ki}=\alpha\gamma_{ki}p_{ki}|c_{ki}|^2
		=\alpha\gamma_{ki}w_{ki},
	\end{equation}
	where $k$ labels the initial state, $i$ the two-meson channel,
	$p_{ki}$ the final-state momentum defined in
	Eq.~\eqref{eq:app_decay_weights},
	and $c_{ki}$ the projection of the normalized color--spin eigenstate
	onto that channel.
	The effective coupling $\alpha$ sets the common normalization,
	while $\gamma_{ki}$ contains the spatial overlap and decay dynamics.
	The spectra are calculated in the diquark--antidiquark color basis
	$\{|\bar33\rangle,|6\bar6\rangle\}$, whereas $c_{ki}$ is the
	singlet--singlet coefficient in the meson-pair basis
	$\{|11\rangle,|88\rangle\}$.
	The color Fierz transformation, spin recoupling, and channel
	projection are given in Eqs.~\eqref{eq:app_decay_color},
	\eqref{eq:app_decay_spin}, and \eqref{eq:app_decay_amplitude},
	respectively, in App.~\ref{app:matrices}.
	For a fixed initial state and approximately equal $\gamma_{ki}$,
	$\Gamma_{ki}/\Gamma_{kj}\simeq w_{ki}/w_{kj}$.
	
	The weights in Table~\ref{tab:fullyheavy14_spectrum_comparison}(a)
	favor pseudoscalar pairs for the lower scalars and vector pairs for
	the upper scalars. In particular, the upper states give
	$\Gamma_{J/\psi J/\psi}/\Gamma_{\eta_c\eta_c}\sim3.7\times10^2$
	and $\Gamma_{\Upsilon\Upsilon}/\Gamma_{\eta_b\eta_b}\sim4.0\times10^2$
	under the same overlap approximation.
	The upper fully charmed scalar can therefore contribute to the
	observed $J/\psi J/\psi$ final state.
	The $1^{+-}$ and $2^{++}$ levels have $S$-wave channels into
	$\eta_Q V_Q$ and $V_QV_Q$, respectively, where
	$(\eta_Q,V_Q)=(\eta_c,J/\psi)$ or $(\eta_b,\Upsilon)$.
	Each level lies above at least one allowed $S$-wave threshold with
	nonzero weight and is thus unstable against strong rearrangement decay.

	\subsection{Hidden-charm hidden-bottom tetraquarks}
	\label{sec:bcbc14}
	
	The $bc\bar b\bar c$ spectrum contains four
	$0^{++}$, two $1^{++}$, four $1^{+-}$, and two $2^{++}$ levels.
	The full spectrum has a spin-averaged mass of 12901 MeV,
	consistent with the trace identities in App.~\ref{app:matrices}.
	Table~\ref{tab:bcbc14_spectrum} summarizes the color-restricted
	and mixed masses, color probabilities, and threshold offsets.
	
	\begin{table*}[ht]
		\caption[Hidden-charm hidden-bottom tetraquark spectrum.]{\label{tab:bcbc14_spectrum}
			\(bc\bar b\bar c\) spectrum and threshold offsets: (a) color-restricted and mixed results; (b) present masses and \(B_c^{(*)}\bar B_c^{(*)}\) threshold offsets. Masses and offsets are in MeV, and probabilities are in percent. The two sets of parentheses after each mass give the prediction scale and the coupling-scan shift, respectively. Parentheses after the threshold offsets give the central \(S\)-wave weights \(w_{ki}\) in Eq.~\eqref{eq:swave_rearrangement_width} (MeV). A dash denotes a forbidden channel, and the superscript \(D\) denotes a \(D\) wave.
		}
		\centering
		\footnotesize
		\setlength{\tabcolsep}{1.8pt}
		\renewcommand{\arraystretch}{1.12}
		\begin{ruledtabular}
			\begin{tabular}{ccrrrrrrrr}
				\multicolumn{10}{c}{
					(a) Color-restricted and mixed masses;
					quarkonium-pair thresholds
				}\\
				$J^{PC}$ & $k$
				& $M_{\bar33}$ & $M_{6\bar6}$
				& $M_{\rm mix}$ & $P_{\bar33}$
				& \shortstack{$\Delta_{\eta_c\eta_b}$\\$(12383)$}
				& \shortstack{$\Delta_{J/\psi\Upsilon}$\\$(12557)$}
				& \shortstack{$\Delta_{\eta_c\Upsilon}$\\$(12445)$}
				& \shortstack{$\Delta_{J/\psi\eta_b}$\\$(12496)$}\\
				\hline
				
				\multirow{4}{*}{$0^{++}$}
				& 1
				& \multirow{4}{*}{
					\begin{tabular}[c]{@{}r@{}}
						$12888(9)({}^{+3}_{-14})$\\
						$12899(9)({}^{+9}_{0})$
				\end{tabular}}
				& \multirow{4}{*}{
					\begin{tabular}[c]{@{}r@{}}
						$12776(9)(3)$\\
						$12895(9)(8)$
				\end{tabular}}
				& $12696(9)({}^{+2}_{-3})$ & 31.840
				& $+314(1174.4)$ & $+139(0.11)$ & --- & ---\\
				& 2 & & &
				$12858(9)({}^{+2}_{-3})$ & 12.790
				& $+475(13.4)$ & $+301(800.6)$ & --- & ---\\
				& 3 & & &
				$12898(9)(6)$ & 99.900
				& $+515(29.2)$ & $+341(368.4)$ & --- & ---\\
				& 4 & & &
				$13006(9)({}^{+1}_{0})$ & 55.470
				& $+623(2.1)$ & $+449(67.5)$ & --- & ---\\
				
				\hline
				\multirow{2}{*}{$1^{++}$}
				& 1
				& \multirow{2}{*}{$12915(9)(5)$}
				& \multirow{2}{*}{$12890(9)(3)$}
				& $12876(9)({}^{0}_{-1})$ & 26.770
				& --- & $+319(1231.3)$ & --- & ---\\
				& 2 & & &
				$12929(9)(3)$ & 73.230
				& --- & $+372(6.7)$ & --- & ---\\
				
				\hline
				\multirow{4}{*}{$1^{+-}$}
				& 1
				& \multirow{4}{*}{
					\begin{tabular}[c]{@{}r@{}}
						$12905(9)({}^{+1}_{-3})$\\
						$12919(9)({}^{+3}_{-1})$
				\end{tabular}}
				& \multirow{4}{*}{
					\begin{tabular}[c]{@{}r@{}}
						$12821(9)(2)$\\
						$12881(9)(2)$
				\end{tabular}}
				& $12762(9)(1)$ & 32.640
				& --- & --- & $+317(1208.4)$ & $+266(0.015)$\\
				& 2 & & &
				$12813(9)(1)$ & 32.350
				& --- & --- & $+368(0.051)$ & $+317(1216.7)$\\
				& 3 & & &
				$12953(9)({}^{+4}_{-3})$ & 77.160
				& --- & --- & $+508(8.9)$ & $+457(11.7)$\\
				& 4 & & &
				$12998(9)(2)$ & 57.850
				& --- & --- & $+554(4.5)$ & $+503(9.5)$\\
				
				\hline
				\multirow{2}{*}{$2^{++}$}
				& 1
				& \multirow{2}{*}{$12953(9)(5)$}
				& \multirow{2}{*}{$12920(9)(3)$}
				& $12876(9)(0)$ & 36.350
				& $+494^{D}$ & $+319(1236.3)$ & --- & ---\\
				& 2 & & &
				$12996(9)({}^{+3}_{-2})$ & 63.650
				& $+614^{D}$ & $+439(1.4)$ & --- & ---\\
			\end{tabular}
		\end{ruledtabular}

		\medskip

		\begin{ruledtabular}
			\begin{tabular}{ccrrrr}
				\multicolumn{6}{c}{
					(b) Present masses and
					$B_c^{(*)}\bar B_c^{(*)}$ threshold offsets
				}\\
				$J^{PC}$ & $k$ & Present
				& \shortstack{$\Delta_{B_c\bar B_c}$\\$(12549)$}
				& \shortstack{$\Delta_{B_c\bar B_c^*}$\\$(12613)$}
				& \shortstack{$\Delta_{B_c^*\bar B_c^*}$\\$(12678)$}\\
				\hline
				
				\multirow{4}{*}{$0^{++}$}
				& 1 & $12696(9)({}^{+2}_{-3})$
				& $+147(96.2)$ & --- & $+18(16.9)$\\
				& 2 & $12858(9)({}^{+2}_{-3})$
				& $+309(791.8)$ & --- & $+180(8.2)$\\
				& 3 & $12898(9)(6)$
				& $+349(480.6)$ & --- & $+220(46.0)$\\
				& 4 & $13006(9)({}^{+1}_{0})$
				& $+457(21.7)$ & --- & $+328(1311.8)$\\
				
				\hline
				\multirow{2}{*}{$1^{++}$}
				& 1 & $12876(9)({}^{0}_{-1})$
				& --- & $+263(206.3)$ & $+198^{D}$\\
				& 2 & $12929(9)(3)$
				& --- & $+316(1193.4)$ & $+251^{D}$\\
				
				\hline
				\multirow{4}{*}{$1^{+-}$}
				& 1 & $12762(9)(1)$
				& --- & $+148(89.4)$ & $+84(23.6)$\\
				& 2 & $12813(9)(1)$
				& --- & $+199(116.2)$ & $+135(26.3)$\\
				& 3 & $12953(9)({}^{+4}_{-3})$
				& --- & $+339(1180.0)$ & $+275(6.3)$\\
				& 4 & $12998(9)(2)$
				& --- & $+385(6.3)$ & $+320(1340.0)$\\
				
				\hline
				\multirow{2}{*}{$2^{++}$}
				& 1 & $12876(9)(0)$
				& $+327^{D}$ & $+263^{D}$ & $+198(103.6)$\\
				& 2 & $12996(9)({}^{+3}_{-2})$
				& $+447^{D}$ & $+383^{D}$ & $+318(1297.5)$\\
			\end{tabular}
		\end{ruledtabular}
	\end{table*}
	
	Color mixing lowers the lightest scalar mass from 12888 MeV
	in the restricted $\bar3_c\otimes3_c$ configuration to 12696 MeV.
	With $P_{\bar33}=31.840\%$, this state is dominated by the
	sextet--antisextet color component.
	The lowest $1^{++}$ and $2^{++}$ states are nearly degenerate
	at about 12876 MeV.
	When the $qq$ couplings are scanned over the range specified above
	while the $q\bar q$ couplings are held fixed,
	the mass shifts of all levels in the full mixed spectrum
	remain within 6 MeV.
	Thus, within the range examined, variations in the chromomagnetic
	couplings have a limited effect on the overall mass scale,
	whereas color mixing plays a particularly important role
	in the low-lying scalar spectrum.

	Table~\ref{tab:bcbc14_models} compares our results with the effective
	diquark spin-interaction model (Diq.) \cite{Xia2026FullyHeavy},
	the nonrelativistic screened-potential diquark model (NRD)
	\cite{Atangana2026Bcbc}, and QCD sum-rule (QCDSR) calculations
	for the scalar and tensor channels
	\cite{Agaev2024BcbcScalar,Agaev2025BcbcTensor}.
	
	\begin{table*}[htbp]
		\caption{\label{tab:bcbc14_models}
			Comparison of $bc\bar b\bar c$ masses with recent
			theoretical results (MeV).
			Diq., NRD, and QCDSR denote the effective diquark
			spin-interaction model, nonrelativistic screened-potential
			diquark model, and QCD sum rules, respectively.
			Levels are ordered independently within each column.
			For our results, the first and second parentheses give
			the prediction scale and coupling-scan shifts, respectively.
			A dash denotes an unreported result.
		}
		\centering
		\small
		\setlength{\tabcolsep}{5pt}
		\renewcommand{\arraystretch}{1.10}
		\begin{ruledtabular}
			\begin{tabular}{crrrrr}
				$J^{PC}$ & Present
				& \shortstack{Diq.\\\cite{Xia2026FullyHeavy}}
				& \shortstack{NRD\\\cite{Atangana2026Bcbc}}
				& \shortstack{QCDSR\\\cite{Agaev2024BcbcScalar}}
				& \shortstack{QCDSR\\\cite{Agaev2025BcbcTensor}}\\
				\hline
				
				$0^{++}$
				& \begin{tabular}[c]{@{}r@{}}
					$12696(9)({}^{+2}_{-3})$\\
					$12858(9)({}^{+2}_{-3})$\\
					$12898(9)(6)$\\
					$13006(9)({}^{+1}_{0})$
				\end{tabular}
				& \begin{tabular}[c]{@{}r@{}}
					12478\\12549\\12633\\12690
				\end{tabular}
				& 12810
				& $12697\pm90$
				& ---\\
				
				\hline
				$1^{++}$
				& \begin{tabular}[c]{@{}r@{}}
					$12876(9)({}^{0}_{-1})$\\
					$12929(9)(3)$
				\end{tabular}
				& \begin{tabular}[c]{@{}r@{}}
					12583\\12666
				\end{tabular}
				& --- & --- & ---\\
				
				\hline
				$1^{+-}$
				& \begin{tabular}[c]{@{}r@{}}
					$12762(9)(1)$\\
					$12813(9)(1)$\\
					$12953(9)({}^{+4}_{-3})$\\
					$12998(9)(2)$
				\end{tabular}
				& \begin{tabular}[c]{@{}r@{}}
					12547\\12591\\12618\\12655
				\end{tabular}
				& 12827 & --- & ---\\
				
				\hline
				$2^{++}$
				& \begin{tabular}[c]{@{}r@{}}
					$12876(9)(0)$\\
					$12996(9)({}^{+3}_{-2})$
				\end{tabular}
				& \begin{tabular}[c]{@{}r@{}}
					12649\\12666
				\end{tabular}
				& 12855
				& ---
				& $12700\pm90$\\
			\end{tabular}
		\end{ruledtabular}
	\end{table*}
	
	Our lowest scalar mass of 12696 MeV agrees closely with
	the sum-rule result of $12697\pm90$ MeV
	in Ref.~\cite{Agaev2024BcbcScalar}.
	In the tensor channel, the screened-potential calculation
	of Ref.~\cite{Atangana2026Bcbc} gives 12855 MeV,
	only 21 MeV below our lower level, placing both predictions
	in a similar mass region.
	The sum-rule tensor estimate of $12700\pm90$ MeV
	\cite{Agaev2025BcbcTensor} is lower,
	indicating some spread among the predictions
	from different approaches.
	
	We apply Eq.~\eqref{eq:swave_rearrangement_width} to the
	$bc\bar b\bar c$ eigenstates using the same recoupling procedure.
	Partial-width ratios are compared separately within the
	quarkonium-pair and $B_c^{(*)}\bar B_c^{(*)}$ families, assuming
	approximately equal spatial overlap factors within each family.
	The weights in Table~\ref{tab:bcbc14_spectrum} favor
	$\eta_c\eta_b$ for the lowest scalar and $J/\psi\Upsilon$ for
	the lower $1^{++}$ and $2^{++}$ levels among quarkonium-pair modes.
	The two lowest $1^{+-}$ levels favor $\eta_c\Upsilon$ and
	$J/\psi\eta_b$, respectively.
	Within the $B_c$-meson family, the lowest scalar gives
	$\Gamma_{B_c\bar B_c}/\Gamma_{B_c^*\bar B_c^*}\simeq5.7$.
	
	The lowest scalar lies about 18 MeV above the
	$B_c^*\bar B_c^*$ threshold and also above the
	$\eta_c\eta_b$ and $B_c\bar B_c$ thresholds;
	the lower tensor lies 319 MeV above the $J/\psi\Upsilon$ threshold.
	All twelve levels therefore lie above at least one open $S$-wave
	threshold.
	
	The LHCb observation of associated prompt
	$J/\psi\Upsilon(1S)$ production
	\cite{LHCb2023JpsiUpsilon}
	provides an experimental basis for searches for the $C$-even
	levels in this final state.
	Angular information could further distinguish the nearly
	degenerate $1^{++}$ and $2^{++}$ contributions.

	\section{Summary and Outlook}
	\label{sec:summary}
	
We have constructed a unified effective mass relation for conventional hadrons and fully heavy tetraquarks, in which a common reduced-mass function and explicit color factors connect the chromoelectric contributions of \(q\bar{q}\) and \(qq\) pairs. The saturating form approaches a finite light-flavor limit while retaining increasingly attractive behavior in the heavy-flavor region. In the common saturation limit, the leading chromoelectric contribution is proportional to the number of constituents and can be absorbed into an effective constituent-mass shift; the residual pair term then accounts for the flavor-dependent deviations.

After subtracting the fixed spin-weighted chromomagnetic contributions, a joint fit to 14 spin-averaged meson and baryon masses yields a root-mean-square residual of about 4~MeV. The eight fitted parameters remain stable when the number of input masses varies from 11 to 17, and the leave-one-out predictive RMS, with \(b\bar{b}\) retained as the heavy-flavor anchor, is about 9~MeV, which we take as a typical empirical uncertainty for the spin-independent masses. With the chromomagnetic interaction included, the same parameter set reproduces the ground-state meson and baryon spectra and gives predictions for doubly and triply heavy baryons consistent with lattice results.

Assuming that the fitted hadron-level constant \(D\) applies to tetraquarks, we diagonalize the sum of the chromoelectric and chromomagnetic mass matrices to obtain the ground-state spectra of \(cc\bar{c}\bar{c}\), \(bb\bar{b}\bar{b}\), and \(bc\bar{b}\bar{c}\). This provides a direct test of extending the conventional-hadron mass baseline to fully heavy tetraquarks. The fully charmed tensor mass is \(6505(9)(9)\)~MeV, consistent with recent lattice determinations, while the lowest \(bc\bar{b}\bar{c}\) scalar mass is \(12696(9)(^{+2}_{-3})\)~MeV. These masses remain relatively stable under variations of the heavy-quark \(qq\) chromomagnetic couplings. Color--spin recoupling gives the relative \(S\)-wave rearrangement-decay weights, which, under the adopted spatial-overlap approximation, identify the favored decay channels. All the obtained levels lie above at least one open \(S\)-wave threshold, so strong rearrangement decays are allowed.

These predictions can be tested by future measurements of the di-\(J/\psi\) invariant-mass spectrum and angular distributions, as well as by studies of the \(\eta_c\eta_c\), \(\Upsilon\Upsilon\), \(J/\psi\Upsilon\), and \(B_c^{(*)}\bar{B}_c^{(*)}\) final states. The light-flavor saturation in this work reflects the model behavior of the effective mass relation at the light-flavor end. When this framework is combined with explicit spatial dynamics, coupled-channel effects should be further incorporated, with particular attention to the mass corrections induced by the coupling of the light-flavor sector to hadronic continua. In addition, the present approach offers a reference for the unified determination of potential-model parameters for conventional and multiquark hadrons.
	
	\begin{acknowledgments}
		
	Wen-Nian Liu thanks Kai Kai Zhang for useful discussions. This work is supported by the National Natural Science Foundation of China (Grant No.~12475026). Support is also provided by the Natural Science Foundation of Gansu Province (Grant No.~25JRRA799).
		
	\end{acknowledgments}
	
	\appendix
	\section{Color--spin bases and spin-independent matrices}
	\label{app:basis}
	
	We specify the color--spin bases and spin-independent matrices
	for $cc\bar c\bar c$, $bb\bar b\bar b$, and $bc\bar b\bar c$,
	including the exchange-symmetry restrictions.

	For the particle order $(1,2,3,4)=(q_1,q_2,\bar q_1,\bar q_2)$,
	let $a,b$ be quark color indices and $c,d$ the conjugate
	antiquark indices. A normalized singlet basis is
	\begin{equation}
		\label{eq:app_color_tensors}
		\begin{aligned}
			(\chi_{\bar33})_{ab;cd}
			&=\frac{\delta_{ac}\delta_{bd}
				-\delta_{ad}\delta_{bc}}{\sqrt{12}},\\
			(\chi_{6\bar6})_{ab;cd}
			&=\frac{\delta_{ac}\delta_{bd}
				+\delta_{ad}\delta_{bc}}{\sqrt{24}}.
		\end{aligned}
	\end{equation}
	These two tensors span the complete color-singlet space.
	They can equivalently be recoupled to meson-pair
	$1_c\otimes1_c$ and $8_c\otimes8_c$ configurations
	\cite{Wu2018FullyHeavy}.

	We denote a coupled spin state by
	$|s_{12}s_{34};JM\rangle$, {where $s_{12}$ and $s_{34}$
		are the pair spins and $J,M$ are the total spin and its projection.}
	We use the standard
	Condon--Shortley convention. In terms of the pair singlet
	$s=(|\uparrow\downarrow\rangle-|\downarrow\uparrow\rangle)/\sqrt2$
	and triplets
	$t_+=|\uparrow\uparrow\rangle$,
	$t_0=(|\uparrow\downarrow\rangle+|\downarrow\uparrow\rangle)/\sqrt2$,
	$t_-=|\downarrow\downarrow\rangle$,
	the relevant states may be fixed by
	\begin{equation}
		\label{eq:app_spin_phases}
		\begin{aligned}
			|00;00\rangle&=s\,s,\\
			|11;00\rangle&=(t_+t_--t_0t_0+t_-t_+)/\sqrt3,\\
			|01;11\rangle&=s\,t_+,\qquad
			|10;11\rangle=t_+\,s,\\
			|11;11\rangle&=(t_+t_0-t_0t_+)/\sqrt2,\\
			|11;22\rangle&=t_+t_+.
		\end{aligned}
	\end{equation}
	Other magnetic components follow by angular-momentum lowering.
	The full basis states are
	$|\alpha;s_{12}s_{34};JM\rangle
	=\chi_\alpha|s_{12}s_{34};JM\rangle$.
	
	Charge conjugation interchanges $(12)$ and $(34)$. With the
	phases above,
	\begin{equation}
		\label{eq:app_charge_conjugation}
		\begin{aligned}
			\mathcal C|\alpha;s_{12}s_{34};JM\rangle
			={}&(-1)^{s_{12}+s_{34}-J}\\
			&\times|\alpha;s_{34}s_{12};JM\rangle.
		\end{aligned}
	\end{equation}
	For each color $\alpha=\bar33,6\bar6$, define
	$|\alpha;\pm\rangle=(|\alpha;01;1\rangle
	\pm|\alpha;10;1\rangle)/\sqrt2$.
	For $bc\bar b\bar c$, set $(q_1,q_2)=(b,c)$.
	Suppressing the magnetic index, the ordered channel bases are
	\begin{equation}
		\label{eq:app_unequal_bases}
		\begin{aligned}
			\mathcal B_{0^{++}}
			&=\{|\alpha;00;0\rangle,|\alpha;11;0\rangle\},\\
			\mathcal B_{1^{++}}
			&=\{|\alpha;+\rangle\},\\
			\mathcal B_{1^{+-}}
			&=\{|\alpha;-\rangle,|\alpha;11;1\rangle\},\\
			\mathcal B_{2^{++}}
			&=\{|\alpha;11;2\rangle\}.
		\end{aligned}
	\end{equation}
	All spin states of $\bar33$ precede those of $6\bar6$.
	The two colors give dimensions $4,2,4,2$, respectively.
	No identical-quark Pauli projection is required for $bc\bar b\bar c$.
	
	For the equal-flavor systems $QQ\bar Q\bar Q$, the bases above
	must also satisfy the Pauli principle.
	For identical quarks in the ground state, the spatial
	and flavor factors are symmetric. The combined color--spin
	factor must therefore be antisymmetric under $1\leftrightarrow2$
	and, separately, $3\leftrightarrow4$.
	The allowed pair combinations are
	$(\bar3_c,s_{12}=1)$ and $(6_c,s_{12}=0)$, with the corresponding
	conjugate antiquark colors and spins. Consequently,
	\begin{equation}
		\label{eq:app_equal_bases}
		\begin{aligned}
			\mathcal B_{0^{++}}^{QQ}
			&=\{|\bar33;11;0\rangle,|6\bar6;00;0\rangle\},\\
			\mathcal B_{1^{+-}}^{QQ}
			&=\{|\bar33;11;1\rangle\},\\
			\mathcal B_{2^{++}}^{QQ}
			&=\{|\bar33;11;2\rangle\}.
		\end{aligned}
	\end{equation}
	There is no $1^{++}$ state in this ground-state basis.
	We impose the Pauli constraints before
	diagonalization and use the bases in Eq.~\eqref{eq:app_equal_bases}.
	The $J=1,2$ states contain $\bar33$ with pair spins $1,1$;
	the $6\bar6$ component has pair spins $0,0$ and contributes to $J=0$.
	
	In the color basis of Eq.~\eqref{eq:app_color_tensors},
	$\mathsf T_{ij}$ represents $\mathbf T_i\!\cdot\!\mathbf T_j$,
	and $\mathbf I_c$ is the identity matrix. The pair-color matrices are
	\begin{equation}
		\label{eq:app_color_matrices}
		\begin{aligned}
			\mathsf T_{12}=\mathsf T_{34}
			&=\begin{pmatrix}-2/3&0\\0&1/3\end{pmatrix},\\
			\mathsf T_{13}=\mathsf T_{24}
			&=\begin{pmatrix}-1/3&-1/\sqrt2\\
				-1/\sqrt2&-5/6\end{pmatrix},\\
			\mathsf T_{14}=\mathsf T_{23}
			&=\begin{pmatrix}-1/3&1/\sqrt2\\
				1/\sqrt2&-5/6\end{pmatrix}.
		\end{aligned}
	\end{equation}
	They satisfy $\sum_{i<j}\mathsf T_{ij}=-(8/3){\mathbf I_c}$.
	Writing $f=F(\mu_{q_1q_2})$ and
	$g=F(m_{q_1}/2)+F(m_{q_2}/2)$ gives
	\begin{equation}
		\label{eq:app_si_matrix}
		\begin{aligned}
			\mathcal M_{\rm SI}
			={}&[2(m_{q_1}+m_{q_2})+D]{\mathbf I_c}\\
			&+\begin{pmatrix}
				-2f-g/3&-(g-2f)/\sqrt2\\
				-(g-2f)/\sqrt2&-f-5g/6
			\end{pmatrix}.
		\end{aligned}
	\end{equation}
	The equal-flavor limit is proportional to the identity,
	consistently with Eq.~\eqref{eq:fullyheavy_baseline14}.
	
	\section{Chromomagnetic matrices and consistency checks}
	\label{app:matrices}
	
	The chromomagnetic matrices below use the bases and phases of
	App.~\ref{app:basis} and enter the full matrix in
	Eq.~\eqref{eq:tetra_full_matrix}.
	
	\subsection{Chromomagnetic matrices}
	\label{app:cmi_matrices}
	
	For $QQ\bar Q\bar Q$ ($Q=c,b$), projecting the chromomagnetic operator onto the Pauli-allowed basis in Eq.~\eqref{eq:app_equal_bases} gives
	\begin{equation}
		\label{eq:app_fullyheavy_cmi}
		\begin{aligned}
			\mathcal H_{\rm CMI}^{0^{++}}
			&=\begin{pmatrix}
				\dfrac{16}{3}(C_{QQ}-2C_{Q\bar Q})
				&-8\sqrt6 C_{Q\bar Q}\\
				-8\sqrt6 C_{Q\bar Q}&8C_{QQ}
			\end{pmatrix},\\
			E_{\rm CMI}^{1^{+-}}
			&=\frac{16}{3}(C_{QQ}-C_{Q\bar Q}),\\
			E_{\rm CMI}^{2^{++}}
			&=\frac{16}{3}(C_{QQ}+C_{Q\bar Q}).
		\end{aligned}
	\end{equation}
	The negative sign of the scalar mixing matrix element originates from the common phase above; reversing the sextet scalar phase yields the equivalent positive-sign convention.
	
	For $bc\bar b\bar c$, all spin states in Eq.~\eqref{eq:app_unequal_bases} are retained. Taking $(1,2,3,4)=(b,c,\bar b,\bar c)$, we define
	\begin{equation}
		\begin{aligned}
			\mathcal{A}&=C_{bc},\qquad \mathcal{B}=C_{b\bar c},\\
			\mathcal{P}&=C_{b\bar b}+C_{c\bar c},\\
			\mathcal{Q}&=C_{b\bar b}-C_{c\bar c},
		\end{aligned}
	\end{equation}
	{
		where these shorthand combinations use equal charge-conjugate chromomagnetic coefficients,
		{$C_{\bar b\bar c}=C_{bc}$ and
			$C_{c\bar b}=C_{b\bar c}$.
			The symbols $\mathcal A,\mathcal B,\mathcal P,\mathcal Q$
			are local energy-valued combinations, distinct from the
			chromoelectric coefficients $A_{\rm col},B_{\rm col}$.}
	}
	For channels containing two spin states, we write
	\begin{equation}
		\label{eq:app_cmi_blocks}
		\mathcal H_{\rm CMI}^{J^{PC}}
		=\begin{pmatrix}
			\mathcal H_{\bar33}^{J^{PC}}&\mathcal V^{J^{PC}}\\
			(\mathcal V^{J^{PC}})^T&\mathcal H_{6\bar6}^{J^{PC}}
		\end{pmatrix},
	\end{equation}
	{where the diagonal blocks retain a single color configuration and its spin mixing, while $\mathcal V^{J^{PC}}$ couples the two color sectors.}
	In the ordered basis of Eq.~\eqref{eq:app_unequal_bases}, the scalar blocks are
	\begin{equation}
		\mathcal H_{\bar33}^{0^{++}}
		=\begin{pmatrix}
			-16\mathcal{A}&-4(\mathcal{P}-2\mathcal{B})/\sqrt3\\
			-4(\mathcal{P}-2\mathcal{B})/\sqrt3&16\mathcal{A}/3-8(\mathcal{P}+2\mathcal{B})/3
		\end{pmatrix},
	\end{equation}
	\begin{equation}
		\mathcal H_{6\bar6}^{0^{++}}
		=\begin{pmatrix}
			8\mathcal{A}&-10(\mathcal{P}-2\mathcal{B})/\sqrt3\\
			-10(\mathcal{P}-2\mathcal{B})/\sqrt3&-8\mathcal{A}/3-20(\mathcal{P}+2\mathcal{B})/3
		\end{pmatrix},
	\end{equation}
	\begin{equation}
		\mathcal V^{0^{++}}
		=\begin{pmatrix}
			0&-2\sqrt6(\mathcal{P}+2\mathcal{B})\\
			-2\sqrt6(\mathcal{P}+2\mathcal{B})&-4\sqrt2(\mathcal{P}-2\mathcal{B})
		\end{pmatrix}.
	\end{equation}
	For $1^{+-}$, the corresponding blocks are
	\begin{equation}
		\mathcal H_{\bar33}^{1^{+-}}
		=\begin{pmatrix}
			-16\mathcal{A}/3-4(\mathcal{P}-2\mathcal{B})/3&-8\mathcal{Q}/3\\
			-8\mathcal{Q}/3&16\mathcal{A}/3-4(\mathcal{P}+2\mathcal{B})/3
		\end{pmatrix},
	\end{equation}
	\begin{equation}
		\mathcal H_{6\bar6}^{1^{+-}}
		=\begin{pmatrix}
			8\mathcal{A}/3-10(\mathcal{P}-2\mathcal{B})/3&-20\mathcal{Q}/3\\
			-20\mathcal{Q}/3&-8\mathcal{A}/3-10(\mathcal{P}+2\mathcal{B})/3
		\end{pmatrix},
	\end{equation}
	\begin{equation}
		\mathcal V^{1^{+-}}
		=\begin{pmatrix}
			-2\sqrt2(\mathcal{P}+2\mathcal{B})&-4\sqrt2 \mathcal{Q}\\
			-4\sqrt2 \mathcal{Q}&-2\sqrt2(\mathcal{P}-2\mathcal{B})
		\end{pmatrix}.
	\end{equation}
	The remaining two channels each have one spin state per color configuration:
	\begin{equation}
		\mathcal H_{\rm CMI}^{1^{++}}
		=\begin{pmatrix}
			-16\mathcal{A}/3+4(\mathcal{P}-2\mathcal{B})/3&2\sqrt2(\mathcal{P}+2\mathcal{B})\\
			2\sqrt2(\mathcal{P}+2\mathcal{B})&8\mathcal{A}/3+10(\mathcal{P}-2\mathcal{B})/3
		\end{pmatrix},
	\end{equation}
	\begin{equation}
		\mathcal H_{\rm CMI}^{2^{++}}
		=\begin{pmatrix}
			16\mathcal{A}/3+4(\mathcal{P}+2\mathcal{B})/3&2\sqrt2(\mathcal{P}-2\mathcal{B})\\
			2\sqrt2(\mathcal{P}-2\mathcal{B})&-8\mathcal{A}/3+10(\mathcal{P}+2\mathcal{B})/3
		\end{pmatrix}.
	\end{equation}
	Combined with Eq.~\eqref{eq:app_si_matrix}, these expressions reproduce the color-restricted and mixed $bc\bar b\bar c$ spectra.
	{
		The relative signs follow from
		Eqs.~\eqref{eq:app_color_tensors} and
		\eqref{eq:app_spin_phases}; the eigenvalues and color probabilities
		are invariant under rephasing of the basis.
	}
	
	\subsection{Spin-averaged masses and numerical checks}
	\label{app:checks}
	
	The matrices above also yield trace identities that relate the spin-averaged spectra to the spin-independent mass matrix. For $bc\bar b\bar c$, the complete color--spin space contains $2\times16=32$ magnetic substates. Since every pair-spin operator has vanishing trace over the full spin space,
	\begin{equation}
		\label{eq:app_centroid_check}
		\frac{\sum_{J^{PC},k}(2J+1)M_{J^{PC},k}}{32}
		=\frac12\operatorname{Tr}_c\mathcal M_{\rm SI}.
	\end{equation}
	{Here $\operatorname{Tr}_c$ denotes the trace over the two color states, and the sum runs over all levels in the listed $J^{PC}$ channels.}
	The resulting {spin-averaged mass} of $bc\bar b\bar c$ is ${12901}$ MeV.
	Each color-restricted spectrum contains 16 magnetic substates,
	{and its spin-averaged mass equals the corresponding diagonal entry of Eq.~\eqref{eq:app_si_matrix}.}
	
	The Pauli-projected $QQ\bar Q\bar Q$ space, by contrast, contains ten substates. Its weighted CMI trace is $56C_{QQ}$, so that
	\begin{equation}
		\label{eq:app_equal_centroid}
		\frac{M_{0^{++},1}+M_{0^{++},2}
			+3M_{1^{+-}}+5M_{2^{++}}}{10}
		=M_{{\rm SI},QQ\bar Q\bar Q}+\frac{28}{5}C_{QQ}.
	\end{equation}
	Equation~\eqref{eq:app_equal_centroid} relates the physical spin average to the spin-independent mass through the term $28C_{QQ}/5$.
	The resulting spectra satisfy these trace identities and the variational ordering condition that the lowest complete-space mass does not exceed either color-restricted minimum in the same $J^{PC}$ channel.
	The spectra are evaluated with full-precision fitted parameters, whereas masses, residuals, and scan shifts are rounded independently.
	Independent constructions reproduce the analytic color--spin matrices and the spin-weighted trace identities.
	
	The channel coefficients $c_{ki}$ in Eq.~\eqref{eq:swave_rearrangement_width} are obtained by recoupling the normalized eigenstates of Eq.~\eqref{eq:tetra_full_matrix} onto color-singlet meson-pair channels. In the color phases of Eq.~\eqref{eq:app_color_tensors},
	\begin{equation}
		\label{eq:app_decay_color}
		\begin{aligned}
			U^{(13)(24)}&=\begin{pmatrix}
				1/\sqrt3&\sqrt{2/3}\\-\sqrt{2/3}&1/\sqrt3
			\end{pmatrix},\\
			U^{(14)(23)}&=\begin{pmatrix}
				-1/\sqrt3&\sqrt{2/3}\\\sqrt{2/3}&1/\sqrt3
			\end{pmatrix},
		\end{aligned}
	\end{equation}
	where the rows correspond to $1_c\otimes1_c$ and $(8_c\otimes8_c)_{1_c}$, and the columns to $\bar3_c\otimes3_c$ and $6_c\otimes\bar6_c$.
	The $(13)(24)$ spin coefficients are
	\begin{equation}
		\label{eq:app_decay_spin}
		\begin{aligned}
			R^J_{s_{13}s_{24};s_{12}s_{34}}
			={}&\widehat s_{12}\widehat s_{34}\\
			&\times\widehat s_{13}\widehat s_{24}
			\begin{Bmatrix}
				\tfrac12&\tfrac12&s_{12}\\
				\tfrac12&\tfrac12&s_{34}\\
				s_{13}&s_{24}&J
			\end{Bmatrix},
		\end{aligned}
	\end{equation}
	with $\widehat s=\sqrt{2s+1}$.
	The $(14)(23)$ coefficients follow by exchanging particles 3 and 4, including the pair-spin phase $(-1)^{1-s_{34}}$.
	For eigenvector coefficients $v_{ab}^{(k)}$, where $a$ and $b$ label color and pair spin, the singlet-channel amplitude is
	\begin{equation}
		\label{eq:app_decay_amplitude}
		c_{ki}=\sum_{a,b}U_{1a}R^J_{ib}v_{ab}^{(k)}.
	\end{equation}
	Meson placements and charge-conjugate modes representing the same physical channel are combined with normalized coefficients and the required $C$ parity, and each physical channel is counted once.
	
	With these projections, the central weights are computed as
	\begin{equation}
		\label{eq:app_decay_weights}
		\begin{aligned}
			p_{ki}&=\frac{\sqrt{\lambda(M_k^2,m_{i1}^2,m_{i2}^2)}}{2M_k},\\
			w_{ki}&=p_{ki}|c_{ki}|^2,
		\end{aligned}
	\end{equation}
	where $M_k=M_k^{\rm cal}$, $m_{i1},m_{i2}$ are the final-state meson masses, and $\lambda(x,y,z)=(x-y-z)^2-4yz$.
	These weights enter the partial-width expression in Eq.~\eqref{eq:swave_rearrangement_width}.
	Only open channels allowed by the conserved quantum numbers are listed, and the quoted weights are approximate central estimates reconstructed from the printed parameters and masses.
	
	\bibliographystyle{apsrev4-1}
	\bibliography{hepref,decay_references_added}
\end{document}